\documentclass{article}
\usepackage{amsmath}
\usepackage{arxiv}
\usepackage[utf8]{inputenc} 
\usepackage[T1]{fontenc}    
\usepackage[hidelinks]{hyperref}
\usepackage{url}            
\usepackage{booktabs}       
\usepackage{amsfonts}       
\usepackage{nicefrac}       
\usepackage{microtype}  
\usepackage{cleveref}       
\usepackage{lipsum}         
\usepackage{graphicx}
\usepackage{doi}
\usepackage{graphicx} 
\usepackage{booktabs}
\usepackage{longtable} 
\usepackage{array} 
\usepackage{url}
\usepackage[numbers,sort&compress]{natbib}
\usepackage{tabularx}
\usepackage{pdflscape}
\usepackage{ragged2e}
\usepackage[table]{xcolor}
\usepackage{enumitem}
\usepackage[UKenglish]{babel}
\usepackage{authblk}
\usepackage{float}
 
\title{Preferentiality and bandwidth drive tie activity in online and offline ego networks}
\date{\today}

\usepackage{authblk}

\author[1]{Gamal Adel}

\author[2]{Shrichand Bhuria}

\author[3,8]{Alessandro Catalano}

\author[4]{Liber Dorizzi}

\author[5]{Leonardo Federici}

\author[6]{Theodora Moldovan}

\author[7, 8]{Berné Nortier}

\author[9]{Chara Deanna Punzal}

\author[10]{Giulia de Meijere\thanks{\texttt{giulia.demeijere@tuni.fi}}}

\author[10,11]{Gerardo Iñiguez\thanks{\texttt{gerardo.iniguez@tuni.fi}}}

\affil[1]{Network Science Group, Leiden Institute of Advanced Computer Science, Leiden University, The Netherlands}
\affil[2]{Inserm U1309, Ar\`enes--UMR 6051, EHESP, CNRS, IEP Rennes, University of Rennes, Rennes, France}
\affil[3]{Department of Physics and Astronomy ``Ettore Majorana'', University of Catania, Catania, Italy}
\affil[4]{IMT School for Advanced Studies, Lucca, Italy}
\affil[5]{Network Science Institute, Northeastern University London, London, United Kingdom}
\affil[6]{InfoLab, Department of Information Technology, Uppsala University, Sweden}
\affil[7]{Department of Computer Science, University of St. Andrews, St. Andrews, Scotland}
\affil[8]{Department of Network and Data Science, Central European University Vienna, Vienna, Austria}
\affil[9]{Data Science Program, College of Science, University of the Philippines Diliman, Philippines}
\affil[10]{Tampere Complexity Lab, Data Science Research Centre, Tampere University, Tampere, Finland}
\affil[11]{Centro de Ciencias de la Complejidad, Universidad Nacional Autonóma de México, Ciudad de México, Mexico}

\begin{document}
\maketitle

\begin{abstract}\small
    Ego networks capture the variety of structural patterns in the social interactions of individuals.
    Recently it has been shown that ego networks in online settings display universal patterns of tie strength distributions, but it is unclear how 
    constraints such as spatial proximity and bounded social bandwidth affect such generic behaviour in offline settings.
    Here, we analyse the time evolution of interaction activity in ego networks constructed from offline face-to-face and colocation data, compare them to online communication networks, and explore simple cumulative advantage models that capture the varying preferentiality of individuals for specific social ties.
    We find that patterns of preferentiality at the population level are similar for online and face-to-face networks, but not for colocation data, suggesting that the latter is a poor proxy of social network structure.
    We also provide evidence that empirical ego networks exhibit a bandwidth in the way communication events are allocated across connections. A model implementing this notion uncovers evidence of universal scaling between the tie preferentiality and bandwidth of individuals, common to all online and offline systems explored.
    Our findings strengthen our understanding of the fundamental mechanisms governing human communication and    
    help disentangle the internal and external factors shaping tie evolution across social contexts.
\end{abstract}

\keywords{social networks \and egocentric networks \and tie strength \and online behaviour \and face-to-face interactions} 

\section{Introduction}
Social networks are central to the spread of ideas, norms, and cultural practices, influencing how people interact, cooperate, and build communities~\citep{wasserman1994social, tomasello2010origins}. Identifying universal patterns in social networks can help us understand how social structure shapes, and is shaped by, individual and collective behaviour~\citep{iniguez2023universal, toivonen2009comparative}.
A defining characteristic of social networks is the variation of tie strength, where strong ties tend to be concentrated within closely connected social circles, and weak ties maintain connectivity and cohesion across the network~\citep{granovetter1973strength}. An ego network consists of the ties connecting an individual (the ego) to their social contacts (the alters). These networks are thus characterized by a small number of strong, close relationships and a larger number of weaker ties. At the individual level, ego networks provide a framework for examining alter selection as a process of social decision-making ~\citep{iniguez2023universal}. 

With the unprecedented growth and adoption of digital communication technologies, many social interactions have moved to online communication channels. A question then arises on the ability of online contexts to display `real' patterns of human contact like those found offline. Some studies highlight similarities between offline and online communication networks, with both sharing high-level structural characteristics, such as degree centrality and average path length, and both driven by homophily~\citep{bastos2021spatializing, kostakos2010making, socievole2015multi}. Online ego networks frequently act as direct replicas of offline familiarity \citep{unlusoy2013learning, arnaboldi2017online}. On the other hand, fundamental differences emerge in the depth, composition, and psychological impact of these ties. Offline relationships are generally more holistic, exhibiting greater multiplexity through a wider variety of shared activities, deeper discussions of intimate topics, and longer durations, which collectively foster stronger emotional bonds \citep{mesch2006quality}. Conversely, online interactions are more heavily dominated by weaker ties, partly due to the removal of geographic barriers, which enables nearly unlimited audience sizes \citep{filiposka2017bridging}. 

Given these differences in tie composition and strength between the two mediums, a more fundamental question arises as to the constraints that govern the selection of alters within an ego network. The characteristic structural pattern of these networks is influenced by humans' ability to maintain social relationships, including limited information processing capacity~\citep{miller1956magical}, social cognition~\citep{bernard1973social, dunbar1998social, dunbar2024social, tamarit2018cognitive, david2013processing}, time availability~\citep{gonccalves2011modeling, miritello2013time, lerman2016information}, and socio-economic and geographical factors. In offline networks, social interaction is further constrained by physical location and the social composition of an individual's environment~\citep{huckfeldt1983social}. Online social media overcomes such constraints by potentially providing long-range interactions. However, as originally proposed by the social brain hypothesis, there is a cognitive constraint on the size of social networks that the communication advantages of online media are also unable to overcome~\citep{dunbar2016online}.

In most social networks, it has been observed that ego networks of most nodes tend to be heterogeneously distributed, where most communications/interactions by the ego are had with a minority of its alters~\citep{toivonen2009comparative,simon1955class}. Previous work has identified factors contributing to this non-uniformity such as preferential attachment~\citep{simon1955class,barabasi1999emergence} and cumulative advantage~\citep{price1976general,barrat2004weighted}. These dynamic mechanisms have been shown to form consistent patterns in ego communication networks~\citep{saramaki2014persistence}. More interestingly, \citet{iniguez2023universal} found universality in tie strength distributions and their individual-level variation across online communication modes. The observed universality arises from the competition between cumulative advantage and random choice, two tie reinforcement mechanisms whose balance determines the diversity of tie strengths.

While these dynamics are observed universally across online communication mediums, it has been shown that individuals have their own social signatures, a tendency to form varying ego network topologies. These social signatures have been identified as being consistent in behaviour both within and across online communication networks~\citep{heydari2024disentangling}. \citet{iniguez2023universal} have shown that egos with specific social signatures tend to maintain that same signature despite large changes in alter composition. This indicates that these signatures are, in part, a characteristic behaviour of the individual~\citep{heydari2018multichannel, heydari2024disentangling}.  

However, while these patterns of ego networks have been well studied in online communication networks, an equivalent analysis on offline networks has not been done. Though many of the constraints that shape heterogeneity of communication apply for both types, there are more factors that could potentially influence the offline ego network topology, such as the limitations of physical space. Whether these additional factors cause a significant deviation in ego network patterns in offline settings compared to online communication networks remains unknown. 

Here, we address this gap by directly comparing the structure of ego networks across 12 online and 16 offline communication datasets. First, we ask whether the social signatures previously documented for online communication hold offline. Second, we show that online and face-to-face networks represent social ties of communication, but colocation networks do not. Third, we demonstrate that online and offline ego networks display a finite bandwidth in the reinforcement mechanism leading to heterogeneous activities with the alters. Fourth, we formalize this intuition by modelling the dynamics using this bandwidth of individuals introducing a parameter, named hereafter as critical activity. Finally, we show evidence of scaling between heterogeneity and bandwidth across online/offline social contexts. 

\section{Results}
\label{sec:results}

We analyse time-stamped pairwise interaction data from two broad settings: online communication and offline physical proximity and interaction. The online data comprise 12 datasets covering mobile-phone short messages, email exchanges, Facebook wall posts, private online messages, forum comments, dating-site interactions, and private messages in a university social network. The offline data comprise 16 datasets measured in hospitals, offices, schools, universities, conferences, and an exhibition. We further separate these offline datasets into two measurement classes: face-to-face contact networks, which capture close-range social interactions, and colocation networks, which capture broader shared presence through radio-frequency identification (RFID), Bluetooth, or Wi-Fi-based measurements.


From each dataset, we construct ego networks by treating every individual as an ego and all individuals with whom the ego has at least one recorded event as alters. To illustrate the temporal structure of these ego networks, we follow the time series of events along the five most active ego--alter connections for a selected ego in one online and one offline dataset (Figure~\ref{fig:fig1}a). These examples show substantial heterogeneity in inter-event times (IETs) and alter activity, defined as the number of events recorded on an ego--alter tie. Activity is concentrated on a small number of alters, with the most active alters accumulating many more events with the ego than the remaining alters. This is visible both in the heterogeneous widths of ties in the aggregated egocentric network and in the faster growth of cumulative activity for some alters over time.

The descriptive statistics reveal three broader patterns across the full dataset collection. First, measurement modality affects the reconstructed ego network scale. Colocation datasets generally produce larger and denser ego networks than face-to-face datasets, with higher mean degree $\langle k\rangle$, mean strength $\langle \tau\rangle$, and maximum alter activity $\langle a_m\rangle$ in many settings (Table~\ref{tab:basic_properties_iet} and Figure~\ref{fig:appendix_summary_groups}a--c). This reflects the fact that colocation, Bluetooth, and Wi-Fi measurements capture broader shared presence, whereas face-to-face sensors capture more restrictive close-range interactions. Second, ego-level activity is heterogeneous across online communication, face-to-face contact, and colocation/proximity networks. The full distributions of degree $k$, strength $\tau$, mean alter activity $t=\tau/k$, minimum alter activity $a_0$, and maximum alter activity $a_m$ span several orders of magnitude (Figure~\ref{fig:basic_properties_three_blocks}), indicating that average values alone do not capture the diversity of ego network activity. Third, repeated ego--alter interactions are temporally bursty across all three measurement classes, as shown by positive burstiness values (Table~\ref{tab:basic_properties_iet} and Figure~\ref{fig:appendix_summary_groups}d) and broad normalized inter-event-time distributions (Figure~\ref{fig:iet_combined}). Together, these descriptive results establish that heterogeneity and burstiness are common features of the data, while the scale and density of reconstructed ego networks depend on the measurement modality. 
Full dataset descriptions are provided in Appendix~\ref{sec:data_code}, with online datasets listed in Table~\ref{tab:online_data_sources} and offline datasets in Table~\ref{tab:offline_data_sources}.

\subsection{Time evolution of social signatures in online and offline contexts}

\begin{figure*}
\centering
\includegraphics[width=\textwidth]{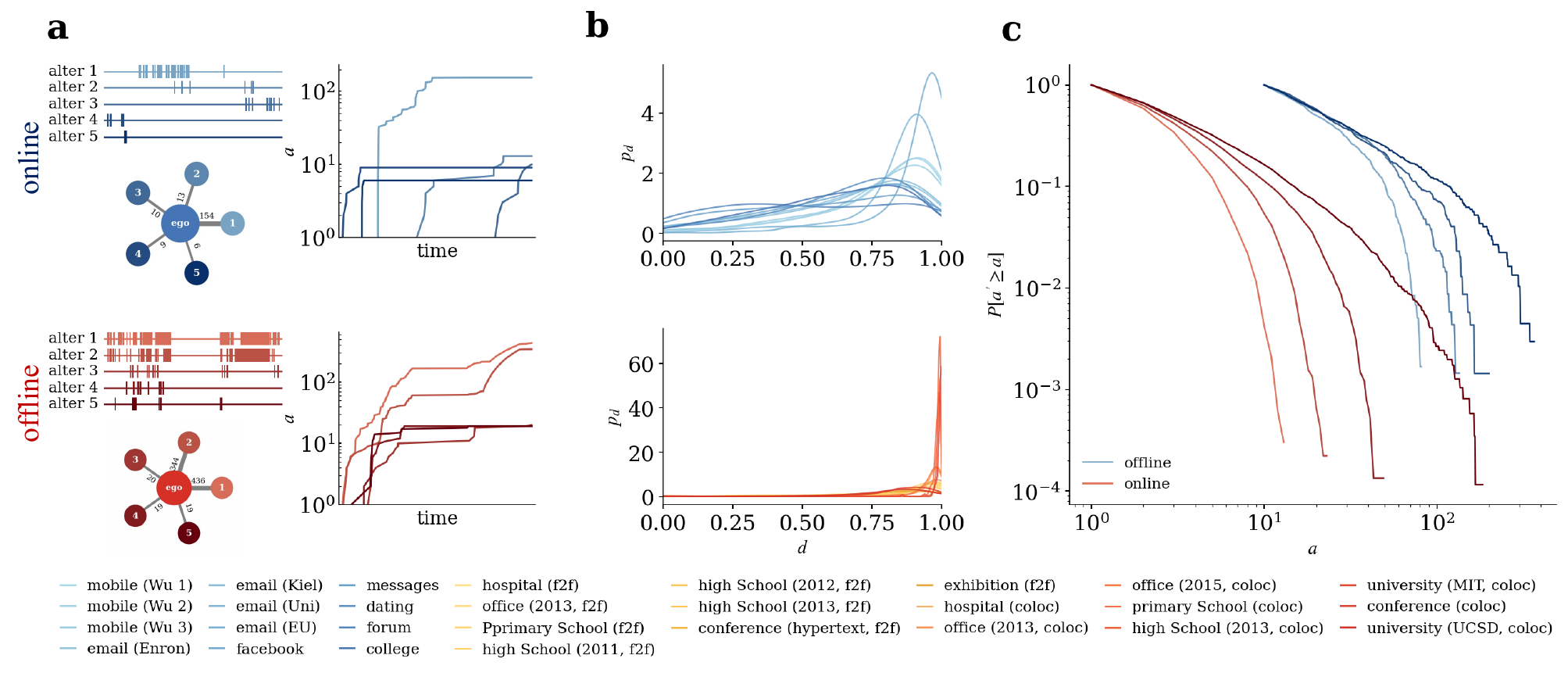}
\caption{\small \textbf{Tie strength across online and offline communication networks.} \textbf{a.} Real-time event sequence between an ego and its $k$ alters and timeline of cumulative activity $a$ (right), for a selected ego in an online channel (College, top) and an offline channel (Primary School (Lyon) face-to-face, bottom). Times are relative to the observation length, so closely spaced events appear as dense lines (left) and sudden increases in $a$ (right). 
With time, some alters accumulate more activity than others. The inset (bottom, leftmost column) shows the aggregated egocentric network, with edge width proportional to tie strength. \textbf{b.} Kernel density estimate (KDE) curves of the dispersion distribution $p_d$ across all datasets, separated into online channels (top, 12 datasets) and offline channels (bottom, 16 datasets), where the dispersion $d$ measures the heterogeneity of alter activities within an ego network. Offline interactions concentrate sharply near $d\approx 1$, indicating heterogeneous tie strengths, whereas online channels span a broader range of dispersion values. \textbf{c.} Complementary cumulative distribution function (CCDF) $P[a' \geq a]$ of the number of alters with activity at least $a$, for egos in each quartile range of the dispersion distribution $p_d$ with $k \geq 10$, shown for College dataset (blue) and Primary School (f2f) (red). Lighter to darker shades denote increasing dispersion quartiles. Broader, heavier-tailed distributions correspond to more heterogeneous social signatures.} 
\label{fig:fig1}
\end{figure*}

Ego signatures are characterized by the dispersion index $d$ of their alter-activity distribution:
\begin{equation}
\label{eq:dispersion_def}
    d = \frac{\sigma^2 - t + a_0}{\sigma^2 + t - a_0},
\end{equation}
where $\sigma^2$ is the variance, $t$ the mean and $a_0$ the minimum of an ego's alter activity.  
Low dispersion ($d\to 0$) marks a homogenous ego that splits its interactions evenly across alters, while high dispersion ($d\to 1$) marks a heterogenous distribution whose activity concentrates on a few dominant ties. Online ego networks span the full range of $d$: most are heterogenous, but a substantial fraction distribute their interactions more evenly, reproducing the broad diversity of social interaction signatures previously reported for online communications \cite{iniguez2023universal} (see Figure~\ref{fig:fig1}b). Offline ego networks, by contrast, collapse onto the heterogenous extreme, with $d$ concentrated near 1. Face-to-face and colocation interactions distribute a small set of dominant alters leaving almost no homogenous tail. The narrowing of the dispersion distribution in offline ego networks reflects how additional constraints of physical presence \cite{huckfeldt1983social} shape ego network structure, compressing the diversity of social communications. 

We use the dispersion and the activity distribution shapes to track how within each dataset, higher-dispersion quartiles produce broader, heavier-tailed complementary cumulative distributions (CCDF). In Figure~\ref{fig:fig1}c, we show that across datasets, the online and offline curves occupy distinct regions of the activity axis: the offline activity distribution systematically extends to higher activity values $a$ than online curves suggesting that offline systems could be associated with more over-dispersed egos. This is consistent with the higher per-tie activity seen in Figure~\ref{fig:fig1}a and Table~\ref{tab:basic_properties_iet}, and the offline CCDFs sitting almost entirely in the high-dispersion regime mirrors the $d\to 1$ accumulation of panel b.  

\begin{figure*} 
\centering
\includegraphics[width=0.80\textwidth]{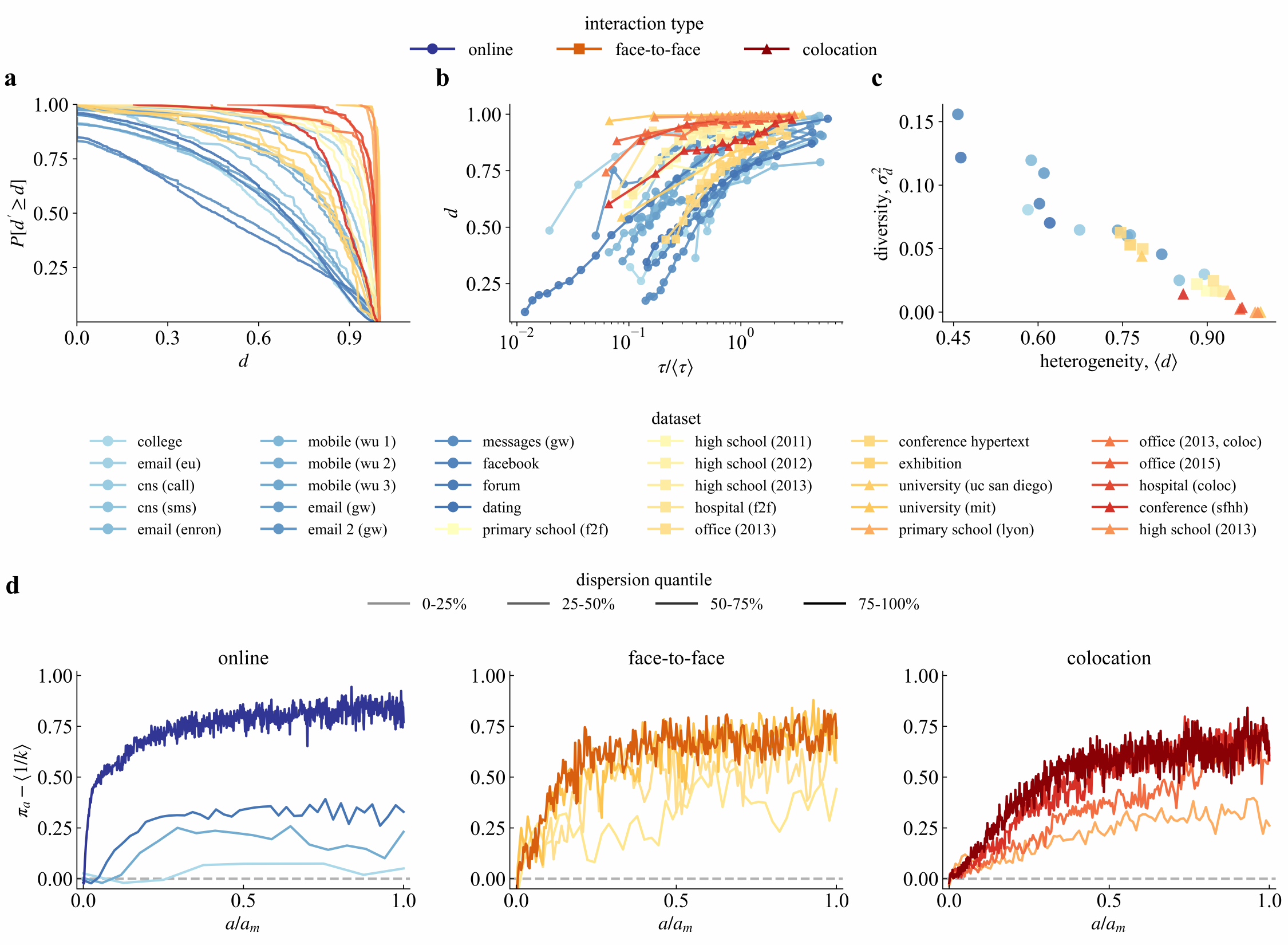}
\caption{\small \textbf{Dispersion and connection kernels of online, face-to-face, ad colocation networks.}
\textbf{a.} CCDFs of dispersion across egos within the different datasets (online-blue, face-to-face (yellow), and colocation (red)).  Online users tend to have more diverse interaction behaviours, while colocation datasets have the least diversity (between egos) and most heterogeneous (within ego) activity behaviours. \textbf{b.} Dispersion $ d $ as a function of strength normalized by the mean values for each dataset $\tau / \langle \tau \rangle$ binned by quantiles of egos. Diversity increases with the strength, but colocation datasets have high values of heterogeneity even for low values of strength. \textbf{c.} Heterogeneity (average  $ d $) vs diversity ($ \sigma^2_d $) for each dataset. The two quantities exhibit an approximately linear negative relationship. Online datasets show low levels of heterogeneity and high diversity compared to offline datasets, in particular colocation, while face-to-face datasets lie in between the two with higher overlap with online ones. 
\textbf{d.} Relative connection kernel $\pi_a - \langle 1/k \rangle$ as a function of normalised alter activity $a/a_m$ in a representative dataset for each context: Email (Kiel) for online (left), High School (2013) for face-to-face (middle) and Primary School for colocation (right). On the x-axis, activity levels are normalised to $[0,1]$ by the maximum activity level $a_m$. The baseline at $0$ denotes the case where communication events are distributed randomly. Lines correspond to the quartile ranges of the dispersion distribution. 
Dispersion is associated with the level of cumulative advantage in the ego network, with the spread between quartile kernels narrowing from online to face-to-face to colocation networks.
For large dispersion, the growth of the connection kernel $\pi_a$ with activity $a$ indicates that alters of high prior activity receive more communication; for low dispersion, communication events happen more at random. 
In all three contexts $\pi_a$ does not grow without limit but bends over and plateaus at high $a$, the signature of a finite bandwidth rather than the linear kernel expected from unbounded preferential attachment.
Individual datasets are shown in Figs.~\ref{fig:each_kernel_online} and~\ref{fig:each_kernel_offline}.
}
\label{fig:fig2}
\end{figure*}

\subsection{Online and face-to-face networks represent social ties of communication, but colocation networks do not} 

Dispersion analysis reveals further differences between online and offline interaction behaviours. 
The CCDFs of dispersion shown in Figure~\ref{fig:fig2}a indicate that the result of Figure~\ref{fig:fig1}a is robust across datasets with egos communicating on online channels exhibiting smaller diversity in behaviour across alters than egos offline. Additionally, we here show that colocation datasets simultaneously display less diverse inter-ego behaviours (narrower distributions) as well as greater heterogeneity (higher mean dispersion) in the distribution of intra-ego tie strengths compared to online egos. 
Face-to-face datasets generally occupy an intermediate position with significant overlap with online datasets. The different behaviours of the three dataset groups are reflected in the shape of their CCDFs: from online datasets to face-to-face and colocation datasets, the decay becomes progressively less smooth: the curves develop flatter profiles and then fall more abruptly in their upper tail. 

Figure \ref{fig:fig2}b further illustrates these differences by showing the relationship between dispersion and normalized ego strength across the three groups of datasets, binned into equal-population bins of 15 egos. A first observation is that lower ego strength is generally associated with a more homogeneous distribution of tie strengths. As ego activity increases, however, these distributions become significantly more heterogeneous, suggesting that a larger interaction budget is allocated increasingly unevenly across alters, thereby reinforcing the distinction between strong and weak ties. The trend holds for all dataset types, but once again online datasets exhibit a much broader range of dispersion values, with a gradual increase from low- to high-strength egos. This variation becomes narrower in face-to-face datasets and is further reduced in colocation settings, where dispersion remains comparatively high even at low ego strength. Thus, while online interaction patterns vary substantially with ego activity, colocation networks appear consistently heterogeneous across the entire strength range. The ranking of dispersion across groups of datasets observed in Figure~\ref{fig:fig2}a is preserved across ego strength classes in Figure~\ref{fig:fig2}b.
To quantify these differences, we distinguish between {\it homogeneity/heterogeneity} of social signatures across an ego's alters and {\it uniformity/diversity} of social signatures across egos. We use homogeneity/heterogeneity to describe the tie strength distribution of a single ego, so that more heterogeneous egos are those for which the tie strengths have greater variation~\cite{iniguez2023universal}. We quantify this through the average dispersion $\langle d \rangle$.
However, egos may share very different or very similar signatures, even when the signatures themselves are quite nuanced. This is what we refer to as the uniformity/diversity of social signatures and is quantified here through the variance of the dispersion distribution $\sigma_d^2$.

To further investigate this low diversity across ego dispersion in offline colocation datasets, in Figure~\ref{fig:fig2}c we report for each dataset the heterogeneity and the diversity. We find a clear negative relationship between the two quantities, with the points approximately following a decreasing linear trend. Moreover, the datasets occupy distinct regions of the parameter space, which leads to an interesting result: online datasets show the highest levels of interaction diversity but comparatively low heterogeneity, whereas colocation datasets are characterized by lower diversity and greater heterogeneity. This result indicates that online datasets contain a larger fraction of egos whose activity is distributed relatively evenly across their acquaintances. As shown in Figure \ref{fig:fig2}b, these more homogeneous allocation patterns are generally associated with lower ego strength. Moving from online to offline datasets, the proportion of egos exhibiting such homogeneous behaviour progressively decreases, becoming very small in colocation groups.
Face-to-face datasets lie between these two extremes, but overlap more with the online datasets, suggesting that face-to-face interaction strategies are more similar to online interaction patterns than to colocation behaviour. Together, these findings suggest an inverse relationship between diversity and heterogeneity, with online and offline datasets occupying distinct regions in this parameter space. Overall, egos display very diverse social signatures in the population with some egos displaying very homogeneous signatures while others exhibit very heterogeneous ones~\citep{iniguez2023universal}. Instead, in offline colocation networks the vast majority of egos tends to display similarly heterogeneous social signatures.

\subsection{Ego networks display finite bandwidth across communication contexts}

To further investigate the mechanisms responsible for link formation in online and offline settings, we now study the connection kernel $\pi_a$, i.e.\ the probability that an alter with activity $a$ at time $\tau$ will communicate with the ego at time $\tau+1$.
Following \citet{iniguez2023universal}, the kernel is estimated empirically for each ego as follows.  
We replay its time-ordered sequence of events and at every step record the current activity of the contacted alter against the activities available among all alters.
Aggregating over all events, the kernel $\pi_a$ is then the empirical probability that an alter with activity $a$ is the one selected for the next interaction, normalized by how often alters at that activity were available to be chosen.
To explicitly investigate the dependence on heterogeneity, we aggregate egos by their dispersion $d$ and report the relative kernel $\pi_a - \langle 1/k \rangle$, where the baseline $\langle 1/k \rangle$ corresponds to alters being contacted uniformly random.
A kernel rising above this baseline signals \emph{cumulative advantage} (active alters continue to attract more activity), whereas a flat kernel near zero is the signature of effectively random choice.

In online communication networks we recover the pattern reported by \citet{iniguez2023universal}: the strength of cumulative advantage is associated with the dispersion of the ego.
Aggregating over the most heterogeneous egos (highest dispersion quartile), the relative kernel increases monotonically with $a$, whereas for the least dispersed egos it flattens and sits close to the random baseline (Figure~\ref{fig:fig2}d, left; all datasets in Figure~\ref{fig:each_kernel_online}).
Cumulative advantage and random choice thus coexist within a channel, and their balance across the dispersion quartiles mirrors the heterogeneity discussed previously.

Extending the analysis to the offline datasets reveals once again a gradient across interaction modes.
Face-to-face networks behave similarly to online ones: the high-dispersion quartiles have a connection kernel resembling cumulative advantage curves while the lower-dispersion quartiles approach the random baseline
(Figure~\ref{fig:fig2}d, middle; all datasets in Figure~\ref{fig:each_kernel_offline}).
Both mechanisms are therefore present, but, consistently with the reduced spread of the dispersion distributions reported above, the quartile kernels are drawn closer together than online, indicating a narrower range of individual behaviour.
Turning to colocation networks, we find this tendency more pronounced, with quartile kernels drawn closer together. Crucially, the dispersion analysis in the top row of Fig~\ref{fig:fig2} shows that their egos are globally shifted toward large $d$, so that all four quartiles correspond to high-dispersion egos.
In this case, their kernels lie well above the random baseline (Fig~\ref{fig:fig2}d, right). Figure~\ref{fig:kernel_all_egos} shows the relative connection kernel for all datasets.

Read together, the kernels provide a mechanistic confirmation of the dispersion results -- the progressive disappearance of homogeneous, random-choice egos from online to face-to-face to colocation is linked to the shift of dispersion distributions toward higher $d$ while simultaneously narrowing across egos. 
Furthermore, in all three contexts, the relative kernel does not grow without limit but flattens at some point, plateauing at high activity. 
We interpret this gradient through the idea of a finite bandwidth, a bound on the effort an ego can distribute across its alters. 
An unbounded preferential attachment mechanism would predict a linear connection kernel~\cite{iniguez2023universal}, a phenomenon which is not observed in the empirical data. 

\subsection{Finite preferentiality generates a hybrid regime of random choice and cumulative advantage}

The results in Figure~\ref{fig:fig2} provide empirical backing to the intuition that individuals do not possess an inexhaustible ability to increase their social interactions with all alters, visible in the flattening of the empirical connection kernel for high activity levels. To capture these plateaus, we 
study a preferential-attachment model of social dynamics 
similar to Price’s model~\cite{price1976general} and adapt it to account for an individual's bandwidth by introducing an additional model parameter of critical capacity.
The probability of communication between an ego and an alter is assumed to depend on their prior communication activity through a parameter $\alpha$ modulating random choice and a critical activity parameter $a_c$. When $a_c\to\infty$, our model reduces to that previously studied by \citep{iniguez2023universal}. 

\begin{figure*} 
\centering
\includegraphics[width=\textwidth]{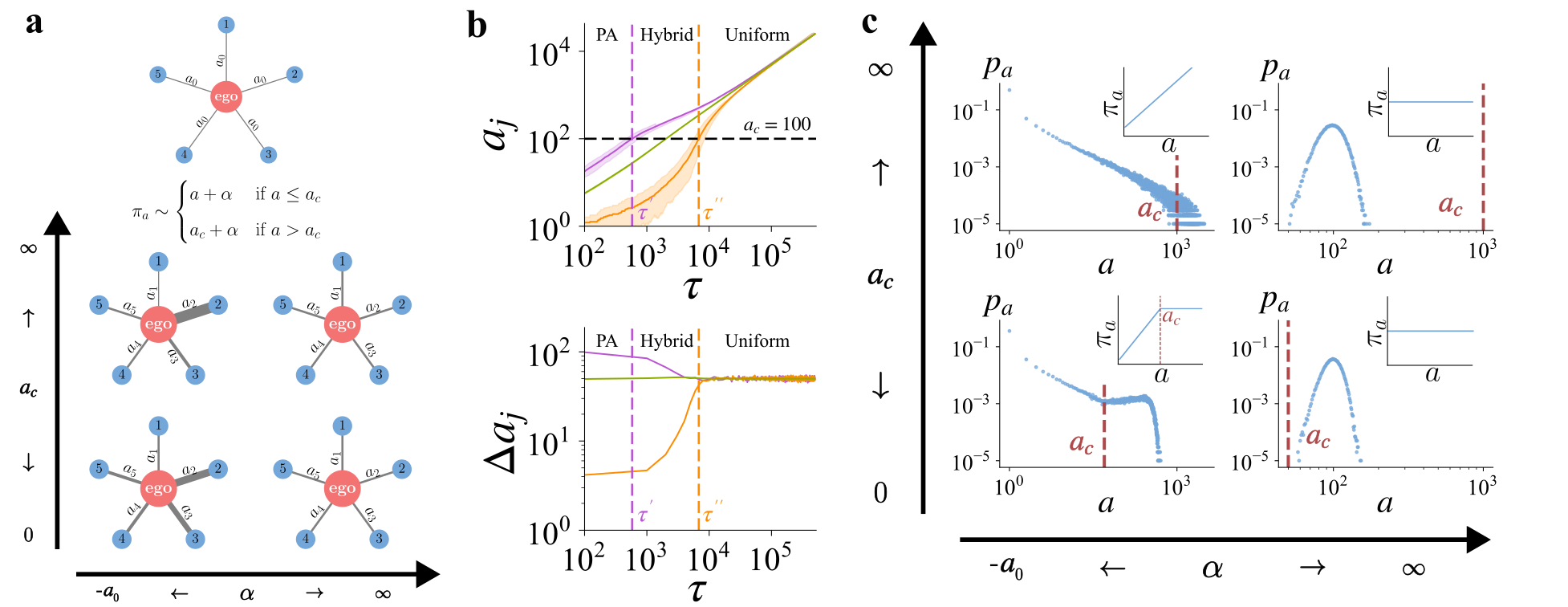}
\caption{\small \textbf{The activity bandwidth model exhibits different dynamical regimes.} 
\textbf{a.} 
For an ego network of degree $k$, alters begin with activity $a_0$ and, at time $\tau$, engage in new communication events with probability $\pi_a$ and critical activity $a_c$. 
The ego networks present tie strength distributions 
typical to each of the 4 different regions in the $(\alpha, a_c)$ phase space, where $\alpha$ tunes the interaction preference while the bandwidth $a_c$ determines when this preferentiality disappears.
Unjoined arrows indicate that the lower-left ego network is not evaluated at exactly $a_c=0, \alpha=-a_0$. \textbf{b.} The bandwidth parameter $a_c$ divides the evolution of alters' tie strengths $\tau_j = \sum_{j}{a_j}$ into 3 distinct regimes, marked by the points at which the first alter (purple) and the last alter (orange) reaches critical activity level $\tau=a_c$. In the PA-regime, one alter (purple) has a high rate $\Delta a_j$ of tie strength change due to preferential attachment. Once critical capacity has been reached, tie strength dynamics enter a hybrid regime marked by a slower rate for the first alter while the less popular alters (green, orange) reach bandwidth one by one. Finally, the last alter (orange) reaches bandwidth and alters are now chosen almost at random, marking the onset of the uniform regime. 
Results are averaged over 50  
samples and in the bottom panel of~\ref{fig:fig3}b, shaded areas indicate 1 standard deviation and we set $k=50, \alpha=0, \text{ and } a_c=100$. \textbf{c.}
The probability distribution $p_a$ that an alter has activity $a$ at time $\tau$ takes different functional forms in the $(\alpha, a_c)$ phase space. For the 4 different limiting behaviours, scatter plots (green) of the $p_a$ distribution and diagrammatic representations in the insets show the connection kernel $\pi_a$.
Here, $k = 50, \tau = 100, \text{ and } \alpha_0=1$. 
}
\label{fig:fig3}
\end{figure*}

The setup proceeds as follows. An undirected ego network with $k$ alters is initialised so that all alters have initial communication activity $a_0$, see Figure~\ref{fig:fig3} top. 
At each time step $\tau$, one alter interacts with the ego, and the probability that an alter with activity $a$ interacts with the ego at time $\tau+1$ is a function $\pi_a$  of both the parameter $\alpha$ and the bandwidth $a_c$ parameters, 
\begin{align}
\label{eq:kernel1}
    \pi_a
    &\sim
    \begin{cases}
        a + \alpha & \text{if } a \leq a_c \\
        a_c + \alpha & \text{if } a > a_c. 
    \end{cases}
\end{align}

The parameter $\alpha$ controls the heterogeneity present in the edge weights and in the limit $\alpha\to\infty$ all alters interact with the ego uniformly and at random (right-hand column, Figure~\ref{fig:fig3}a).
Conversely, when $\alpha$ is small the edge weights become increasingly heterogeneous and the shape of the probability distribution $p_a$ ( the probability that an alter with activity $a$ at time $\tau$ will interact with the ego)  is controlled by the critical capacity parameter $a_c$ (left-hand column, Figure~\ref{fig:fig3}a), separating the $(\alpha, a_c)$ phase space into distinct regions. 
It is important to note that an alter reaching bandwidth does not imply that it {\it stops} engaging in communication activity, and merely that the probability of interaction is no longer a linear function of its current activity level $a$. As such, it will continue engaging in activity with the ego but at a reduced rate. For a more detailed description of the model, see Section~\ref{sec:methods}.


We now track the tie strength $a_j$ as time evolves in Figure~\ref{fig:fig3}b and plot the curves of the alter that reaches $a_c$ first (in purple), last (in orange), and average the remaining alters (in green). Denote by $\tau'$ and $\tau''$ the first- and last-crossover times, respectively. Three different regimes emerge which are separated by $\tau'$ and $\tau''$ -- we call these the Preferential attachment (PA), Hybrid and Uniform regimes.
Initially, an ego is in the PA regime: the communication mechanism is preferential, and some alter benefits from the cumulative advantage to emerge as leader and engage in a very high rate of communication activity with the ego. Figure~\ref{fig:fig3} plots the rate of change in activity $\Delta a_j$ and we observe the curves in green and orange have a very slow rate of increase in total activity compared to the leading alter. Once this alter (purple) reaches critical activity, its connection probability is fixed to $\pi_a=a_c+\alpha$, which in turn allows other alters to benefit from the preferential advantage mechanism. After $\tau'$, the hybrid-PA regime is entered, and the primary alter has a decreasing rate of activity accumulation (negative curve in b, Hybrid regime) while the other alters (green) gradually each reach critical activity threshold. In turn, this allows the last alter (orange) to increase its rate of communication and ultimately also reaches critical activity level at $\tau''$, indicating the end of the Hybrid regime. Once the entire ego network has reached critical capacity, alters {\it continue} engaging in communication activity with the ego, but now communication with the ego occurs with the same probability $\pi_a = a_c + \alpha$ across all alters, i.e. interactions occur uniformly and at random. In Figure~\ref{fig:fig3}b (bottom) this is observable as a roughly equal rate of change in the uniform regime.

Eventually, the probability distribution $p_a$ attains stationary state, which we visualise in Figure~\ref{fig:fig3}c for different regions of the $(\alpha, a_c)$ phase space. The model thus suggests that the existence of a bandwidth in the activity kernels eventually leads to random choice of alters for all egos. The results are reported in Figure~\ref{fig:fig3}c and we plot the respective connection kernels in the panel insets. 
As expected, the uniform connection kernel induced by large $\alpha$ leads to increasingly narrow activity distributions regardless of the value of $a_c$, which indicates that choice becomes effectively random (Figure~\ref{fig:fig3}c, right-hand column). This corresponds to the uniform regime identified in Figure~\ref{fig:fig3}b.
For small $\alpha$, on the other hand, heterogeneity in the activity distribution is increasingly possible. When the bandwidth parameter $a_c$ is large (Figure~\ref{fig:fig3}c top left), the activity distribution $p_a$ takes the broad shape of a gamma distribution, recovering results from the previous, simpler, model 
where connectivity is driven by preferential attachment and alters with high activity at time $\tau$ benefit have higher probability of being selected at activity time $\tau + 1$. Finally, when both $\alpha$ and $a_c$ are small, the novel Hybrid-PA phenomenon emerges, where at first individuals are selected according to a cumulative advantage mechanism but, for $a>a_c$, lose their advantage with an increasing rate. 

\subsection{Evidence of universal scaling between preferentiality and bandwidth across online/offline social contexts} 

\begin{figure*} 
\centering
\includegraphics[width=0.8\textwidth]{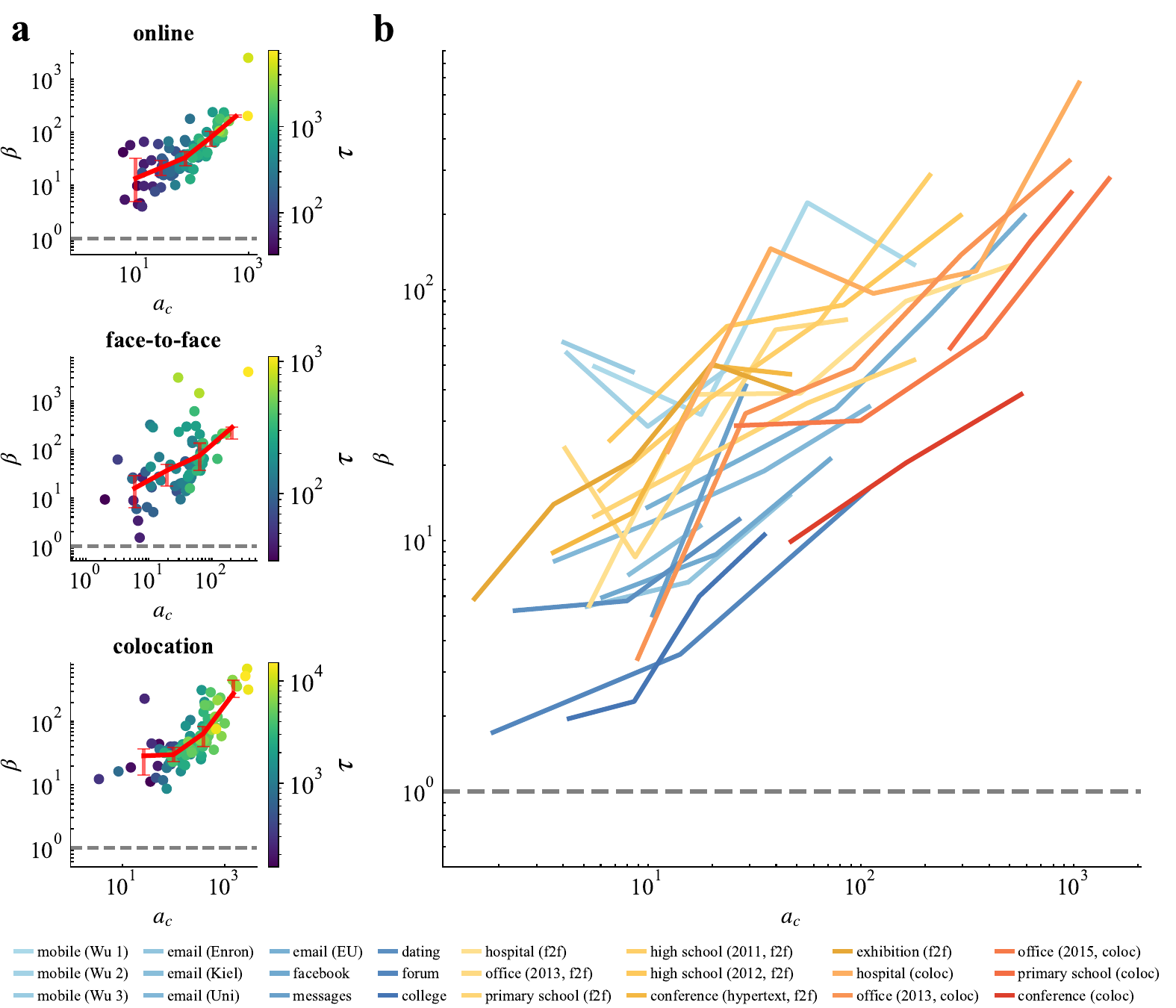}
\caption{\small \textbf{Preferentiality scales with social bandwidth across online and offline contexts.}
For each empirical ego network with strength $\tau \geq 30$, we fit the bandwidth model to obtain associated preferentiality parameter $\beta = t_r/\alpha_r$ and its critical capacity $a_c$ (see Methods~\ref{subsec:fitmodel}); larger $\beta$ corresponds to more heterogeneous tie strengths, with the crossover $\beta = 1$ separating heterogeneous ($\beta > 1$) from homogeneous ($\beta < 1$) ego networks.
\textbf{a.} Fitted preferentiality $\beta$ versus fitted critical activity $a_c$ for every ego in one representative dataset from each measurement group: online (Email (EU), top), face-to-face (High School 2011, middle), and colocation (Office 2015, bottom). Each point is an ego, coloured by strength $\tau$; the red curve traces the binned geometric mean of $\beta$ and the whiskers span the 25th--75th percentiles within each bandwidth bin. The dashed line marks the crossover $\beta = 1$.
\textbf{b.} Binned-mean curves of $\beta$ versus $a_c$ for all datasets, coloured by measurement group (online in blue, face-to-face in yellow, colocation in red).
Across online, face-to-face, and colocation networks, we observe preferentiality and critical activity positively scaling together, where egos with larger social bandwidth also exhibit more heterogeneous tie strengths. With almost all egos lying in the heterogeneous regime ($\beta > 1$), the three classes occupy distinct but overlapping bands, with colocation curves shifted toward the largest capacities.
}
\label{fig:fig4}
\end{figure*}

The empirical kernels of Fig~\ref{fig:fig2}d point beyond a linear mechanism: in every channel, $\pi_a$ rises with activity, signalling cumulative advantage, but saturates at high $a$ rather than growing without bound. This saturation is a kernel-level sign of a finite social bandwidth. It is consistent with a trade-off between diversity and bandwidth, in which, at sufficiently high communication rates, the marginal utility of further reinforcing an already-strong tie, in terms of novel information, diminishes relative to the cost of sustaining that interaction~\citep{aral2011diversity}. The same bound is visible directly in communication data, where the top-ranked alter receives at most $\sim$25\% of an ego's activity and the top three alters about 50\%~\citep{saramaki2014persistence}. The linear-clipped kernel of Eq.~\eqref{eq:kernel1} is the minimal form that captures this behaviour.

To test this picture quantitatively, we fit the model kernel directly to the empirically measured kernel of each ego. We do so in two stages. First, we estimate the random choice parameter $\alpha$ exactly as in the linear model of \citet{iniguez2023universal}, via maximum likelihood on the ego's activity distribution (Section~\ref{subsec:fitmodel}; for fitted values of $\alpha$ in all datasets see Table \ref{tab:table_stats_100}); this also fixes the preferentiality parameter $\beta = t_r/\alpha_r$, which depends only on $\alpha$ and the ego's mean and minimum activity. Second, holding $\alpha$ fixed, we fit the bandwidth $a_c$ by minimising the root mean-squared error between the model kernel $\pi_a = (\min(a,a_c)+\alpha)/\kappa$ and the empirical kernel obtained by replaying the ego's time-ordered event sequence. 
We restrict the analysis to egos with strength $\tau \geq 30$, for which the empirical kernel is sufficiently sampled. Crucially, $\beta$ (preferentiality) and $a_c$ (critical activity) are extracted from independent features of the data (the shape of the activity distribution and the shape of the connection kernel, respectively) so any relationship between them is an empirical finding rather than a built-in constraint.

Figure~\ref{fig:fig4} reveals such a relationship: across both online and offline settings, the fitted preferentiality $\beta$ increases systematically with the fitted bandwidth $a_c$. Egos with a larger social bandwidth thus also tend to have more heterogeneous tie strengths, and the per-ego scatter within a representative dataset for each measurement group (Figure~\ref{fig:fig4}a) confirms that this is a within-channel trend and not an artefact of dataset-level averaging. The binned-mean curves (Figure~\ref{fig:fig4}b) show the same positive scaling for online, face-to-face, and colocation networks alike, with the three classes tracing distinct but overlapping bands and colocation egos shifted toward the largest bandwidths. Almost all egos sit above the crossover $\beta = 1$, placing them firmly in the heterogeneous regime. That heterogeneity and bandwidth co-vary in the same direction regardless of communication medium suggests a common organising principle: the same finite-capacity mechanism that concentrates activity on a few alters also sets how much total interaction an ego can sustain, online and offline alike.

\begin{table}[H]
  \centering
  \small
  \begin{tabular}{l r | r r r r}
    \toprule
    Dataset & $N$ & $n_D$ & $n_{W^2}$ & $n_{U^2}$ & $n_{A^2}$ \\
    \midrule
    Mobile (Wu 1)            & 100 & 0.48 & 0.65 & 0.74 & 0.49 \\
    Mobile (Wu 2)            & 100 & 0.45 & 0.62 & 0.75 & 0.44 \\
    Mobile (Wu 3)            & 100 & 0.33 & 0.49 & 0.60 & 0.32 \\
    Email (Enron)            & 100 & 0.49 & 0.58 & 0.62 & 0.48 \\
    Email (Kiel)             & 100 & 0.34 & 0.40 & 0.47 & 0.34 \\
    Email (Uni)              & 100 & 0.70 & 0.73 & 0.72 & 0.65 \\
    Email (EU)               & 100 & 0.71 & 0.63 & 0.61 & 0.68 \\
    Facebook                 & 100 & 0.65 & 0.69 & 0.64 & 0.68 \\
    Messages                 & 100 & 0.41 & 0.47 & 0.44 & 0.40 \\
    Dating                   & 100 & 0.51 & 0.59 & 0.67 & 0.52 \\
    Forum                    & 100 & 0.49 & 0.49 & 0.50 & 0.49 \\
    College                  & 100 & 0.69 & 0.71 & 0.74 & 0.70 \\
    Hospital (f2f)           &  70 & 0.79 & 0.83 & 0.80 & 0.80 \\
    Office 2013 (f2f)        &  58 & 0.60 & 0.71 & 0.88 & 0.57 \\
    Primary School (f2f)     & 100 & 0.87 & 0.92 & 0.88 & 0.86 \\
    High School 2011         &  75 & 0.71 & 0.85 & 0.85 & 0.69 \\
    High School 2012         & 100 & 0.89 & 0.96 & 0.94 & 0.84 \\
    High School 2013 (f2f)   & 100 & 0.61 & 0.86 & 0.84 & 0.63 \\
    Exhibition               & 100 & 0.42 & 0.72 & 0.85 & 0.43 \\
    Conference (Hypertext)   &  68 & 0.63 & 0.71 & 0.74 & 0.62 \\
    Hospital (coloc)         &  60 & 0.85 & 0.93 & 0.83 & 0.90 \\
    Office 2013 (coloc)      &  91 & 0.65 & 0.64 & 0.45 & 0.68 \\
    Office 2015              & 100 & 0.54 & 0.51 & 0.50 & 0.51 \\
    Primary School (coloc)   & 100 & 0.01 & 0.02 & 0.01 & 0.01 \\
    High School 2013 (coloc) & 100 & 0.31 & 0.30 & 0.27 & 0.31 \\
    Conference (SFHH)        & 100 & 0.58 & 0.55 & 0.54 & 0.50 \\
    \bottomrule
  \end{tabular}
  \vspace{0.8em}
  \caption{\small \textbf{Statistical significance of maximum likelihood estimation.} Fraction $n_\bullet$ of ego networks satisfying the condition $p_\bullet > 0.1$ on the $p$-value $p_\bullet$ associated to the test statistics of Kolmogorov-Smirnov, Cram\'er-von Mises, Watson, and Anderson-Darling [$\bullet = D, W^2, U^2, A^2$, respectively].
  Fractions $n_\bullet$ are calculated relative to the number $N$ of egos in each dataset under the condition $t > a_0$ (i.e.\ with any level of heterogeneity on their communication signatures). The model is able to reproduce observed data for most egos and in all settings, with face-to-face ego networks achieving the best results.}
  \label{tab:table_stats_100}
\end{table}

\section{Discussion}
\label{sec:disc}

Our results show that the heterogeneous, cumulative advantage-driven structure of tie strength previously documented in online communication networks also characterizes face-to-face interaction, but not colocation. This convergence between online and face-to-face networks is notable given how different these mediums are in their constraints: online platforms remove geographic and temporal barriers entirely, while face-to-face interaction is bound by physical co-presence and a finite pool of available others. That a common signature of tie strength dispersion emerges despite this contrast suggests the underlying driver is not the communication channel itself, but a shared social selection process: individuals concentrating attention on a small subset of alters regardless of how that attention is delivered. Colocation networks break this pattern, distributing activity far more evenly across contacts. We interpret this difference as evidence that colocation captures exposure rather than choice: people are routinely co-present with others without exercising any social preference over them, whereas online and face-to-face conversation both require an active decision to engage.

This distinction matters because colocation is frequently used as a proxy for social ties, inferred from Wi-Fi/Bluetooth proximity, cell-tower data, or GPS traces. Direct observation of social interaction is costly or infeasible at scale; thus passive location data is comparatively easy to collect \citep{eagle2006reality, eagle2009inferring, cranshaw2010bridging, sapiezynski2019interaction}. Our results suggest this substitution should be made with caution: colocation networks have a fundamentally different structure from communication networks and do not capture social communication behaviour. They may still be valuable for other purposes, for instance, as a proxy for airborne disease transmission pathways, where physical co-presence, not social closeness, is what matters. In that sense, when it comes to epidemic spreading, we are all homogeneous regardless of how selective we are in our actual social ties.

Individuals are taking decisions continuously about whom to engage with; despite constraints, social signatures are important. The cumulative advantage mechanism driving tie strength dispersion can be understood as people concentrating attention on their closest relationships. These relationships form, in part, from similarities across sociodemographic, behavioural, and intrapersonal characteristics \citep{mcpherson2001birds}. Although the reasons for these similarities differ across mediums (online, people can in principle choose to interact with anyone; offline, school-age children are largely grouped with peers of similar sociodemographic background), the underlying process of alter selection is comparable. This convergence likely reflects a shared selection mechanism rather than a shared medium: the kernel comparisons in Figure~\ref{fig:fig2}d show that face-to-face ties retain the same cumulative advantage shape as online ties, differing mainly in the narrower range of individual dispersion across egos (Figure~\ref{fig:fig2}a--c) rather than in the underlying process itself. Spatial constraints appear to limit how varied an ego's signature can be, without changing the type of mechanism generating it. Once a tie requires active selection rather than passive exposure, the same finite-bandwidth dynamic governs it, whether that selection happens face-to-face or through a screen. This can be explained by the relative stability of cognitive, temporal, and other constraints \citep{miller1956magical, bernard1973social, dunbar1998social, tamarit2018cognitive, gonccalves2011modeling, miritello2013limited}, as well as by individual personality traits \citep{swickert2002extraversion, costa1992four, centellegher2017personality}.

\citet{iniguez2023universal} attribute the universal pattern of heterogeneous tie strengths in online networks and the distribution of the preferentiality parameter to Dunbar's number and tie strength reinforcement. The fact that this same pattern holds for face-to-face communication suggests that these mechanisms may be medium-independent: our bandwidth for social bonding appears to consistently shape network structure regardless of the channel used \citep{dunbar1998social, kock2005media}. A complementary explanation is that the ease of repeated interaction with the same alters itself reinforces tie strength over time. Crucially, both online and face-to-face ties involve a degree of emotional closeness that colocation ties lack, which may be the key factor distinguishing why the two interaction mediums converge on a shared pattern while colocation networks do not.

The empirical data demonstrates the existence of a well-defined social bandwidth within the dynamic process underlying the temporal evolution of ego interactions. In both online and offline environments, the reinforcement mechanisms responsible for generating highly heterogeneous activity distributions among alters do not scale indefinitely. Instead, upon reaching a specific threshold, which we identify as the critical activity, a structural shift occurs: egos begin to homogenize their communicative choices, systematically exploring and re-engaging with alters with whom they had previously maintained only marginal interactions. This behaviour leads to a scaling law between preferentiality and critical activity, which reveals that communication patterns become less heterogeneous when the bandwidth in the connection kernel is lower. The sooner an individual reaches this critical threshold, the earlier and more intensively they will begin to distribute their communicative attention toward a broader, more balanced set of social relationships.

\section{Materials and Methods} 
\label{sec:methods}

\subsection{Data description and processing}

\label{subsec:data_preprocessing}

\paragraph{Empirical data.} The empirical analysis is based on 28 recurring, time-stamped pairwise interaction datasets selected to compare ego network structure across different forms of human interaction. We distinguish between online communication datasets, where events correspond to active exchanges initiated between individuals, and offline physical-proximity/contact datasets, where events correspond to observed face-to-face contact, close-range proximity, or broader co-presence in bounded social settings. 

The 12 online datasets cover several communication contexts and observation timescales, including short messages from three mobile-phone datasets collected over one month~\citep{wu2010evidence}; email communication from the Enron corporation~\citep{klimt2004enron,kunegis2013konect}, Kiel University~\citep{ebel2002scale,saramaki2015exploring}, an unnamed university~\citep{saramaki2015exploring,eckmann2004entropy}, and a large European research institution~\citep{leskovec2007graph,paranjape2017motifs}; Facebook wall-post interactions~\citep{saramaki2015exploring,viswanath2009evolution}; private messages and forum comments from the Swedish movie recommendation site Filmtipset~\citep{saramaki2015exploring,said2010social,karimi2014structural}; multi-modal interactions from the Swedish dating site pussokram.com, including emails, guest-book signings, friendship requests, and friendships~\citep{saramaki2015exploring,holme2004structure}; and private messages from a Facebook-like university social network~\citep{opsahl2009clustering,panzarasa2009patterns}. 

The 16 offline datasets cover physical interaction in hospitals, offices, primary and high schools, universities, conferences, and an exhibition. Most were collected using wearable radio-frequency identification (RFID) sensors following the SocioPatterns protocol, which records close-range proximity with a temporal resolution of 20 seconds and an approximate spatial range of 1--1.5 meters~\citep{genois2018can}. These RFID-based datasets include hospital interactions~\citep{vanhems2013estimating}, office contacts~\citep{genois2018can}, primary-school contacts~\citep{stehle2011high}, high-school contacts~\citep{fournet2014contact,mastrandrea2015contact}, conference contacts~\citep{genois2018can,isella2011s}, and exhibition contacts~\citep{isella2011s}. We also include broader proximity data inferred from Bluetooth handshakes among university students at MIT~\citep{eagle2006reality} and Wi-Fi-based proximity among freshmen at UC San Diego~\citep{mcnett2005access,chaintreau2007impact}. Because offline measurements capture different levels of physical interaction, we classify them as either face-to-face contact networks, representing close-range social interaction, or colocation/proximity networks, representing broader shared presence measured through RFID, Bluetooth, or Wi-Fi-based systems. Detailed descriptions of each dataset, including observation windows, measurement protocols, and source references, are provided in Appendix~\ref{sec:data_code}, with online datasets listed in Table~\ref{tab:online_data_sources} and offline datasets listed in Table~\ref{tab:offline_data_sources}.

\paragraph{Ego network construction.}
\label{subsec:ego_network_construction}
Each dataset is represented as a sequence of time-stamped pairwise events $(u,v,\tau)$, where $u$ and $v$ denote the two interacting individuals and $\tau$ is the event timestamp. We remove self-events, for which $u=v$, and exact duplicate event records. Because the ego network analyses focus on the distribution of activity across an ego's alters, all pairwise events are treated symmetrically: an event between individuals $u$ and $v$ contributes to the ego network of $u$ with alter $v$, and to the ego network of $v$ with alter $u$.

For each dataset, every individual is considered in turn as an ego, and all individuals with whom the ego has at least one recorded event are treated as alters. For each ego--alter pair, we define the alter activity $a$ as the number of events observed on that tie during the observation window. We then compute five ego network activity variables. The degree $k$ is the number of alters of the ego, the strength $\tau$ is the total number of events involving the ego, and the mean alter activity is
$ t = \frac{\tau}{k}.$
The minimum and maximum alter activities are denoted by $a_0$ and $a_m$, respectively.

We retain only egos satisfying
\begin{equation}
    t > a_0.
\end{equation}
This filter removes trivial ego networks in which all alters have identical activity, i.e., $t=a_0$, and ensures that the retained ego networks contain meaningful variation in alter activity. Dataset-level summary statistics for the retained ego networks are reported in Table~\ref{tab:basic_properties_iet}. To compare ego network activity distributions across datasets and measurement classes, we use complementary cumulative distribution functions (CCDFs) of $k$, $\tau$, $t$, $a_0$, and $a_m$.

\paragraph{Edge-level inter-event times and burstiness.}
\label{subsec:iet_burstiness}

Temporal irregularity is measured using edge-level inter-event times. For each ego--alter pair with repeated events, we sort the timestamps and compute the time gaps between consecutive events. To reduce boundary effects from the finite observation window, we discard residual waiting times. The remaining inter-event times are then pooled across all ego--alter pairs within each dataset.
Because datasets differ in timestamp units, temporal resolution, and observation length, raw mean inter-event times $\langle \mathrm{IET}\rangle$ are reported in the original timestamp units and should be interpreted within each dataset rather than compared directly across datasets. For distributional comparison, each inter-event time is normalised by the dataset-specific mean,
\begin{equation}
    X_{\mathrm{IET}} = \frac{\mathrm{IET}}{\langle \mathrm{IET}\rangle}.
\end{equation}
We compare the normalised inter-event-time distributions using CCDFs, $P[X_{\mathrm{IET}}\geq x]$.
We summarize temporal irregularity using the burstiness index
\begin{equation}
    B =
    \frac{\sigma_{\mathrm{IET}}-\langle \mathrm{IET}\rangle}
    {\sigma_{\mathrm{IET}}+\langle \mathrm{IET}\rangle},
\end{equation}
where $\sigma_{\mathrm{IET}}$ is the standard deviation of the pooled edge-level inter-event times. Values of $B>0$ indicate bursty temporal dynamics, where repeated interactions are clustered in time and separated by longer inactive periods. Values close to zero indicate approximately Poisson-like timing, whereas negative values indicate more regular spacing.

\subsection{Model of alter choice with social bandwidth}
\label{subsec:model}

In this section, we introduce a modelling framework for ego network dynamics aimed at capturing the fundamental mechanisms of human social interactions. Consider a network (the so-called {\it ego network}) centred on a single individual (\textit{the ego}) and connected to $k$ neighbours (\textit{alters}). We represent each social tie by a weighted edge, where weight quantifies the cumulative communication activity between the ego and a specific alter. In our work we consider both online (e.g. calls and messages) and offline (e.g. face-to-face, colocation) interactions.
    
The evolution of tie strengths unfolds over discrete event times, starting from $\tau_0$ up to the maximum observation time $T$. We initialise the network homogeneously such that all $k$ alters have baseline activity $a_0 \ge 0$, the initial edge weight. 
At each subsequent time step $\tau+1$, an alter $j$ is chosen with some probability $\pi_a$ and its activity is increased by 1, i.e., $a_j(\tau + 1) = a_j(\tau) + 1$.
By setting the initial event time to $\tau_0 = k a_0$, we ensure that the temporal variable $\tau$ is equal to the total activity of the network up to that time, i.e., satisfying $\tau = \sum_{j}^k a_j$. Consequently, at time $\tau$, the activity of all alter are bounded within the interval $[a_0, \tau]$.

The selection of which alter interacts with the ego at each step is a stochastic process governed by a probability $\pi_a$. This quantity, referred to as the connection kernel, accounts for the history of past interactions. Depending on how behavioural mechanisms are integrated into the model, different functional forms can be adopted. The most straightforward choice assumes that the connection kernel depends linearly on the cumulative past activity $a$ \cite{iniguez2023universal}, embodying a rich-get-richer dynamics \cite{price1976general}. In particular, this implies that a higher frequency of past interactions progressively increases the likelihood of future engagements with the same alter. Within this formulation, the introduction of a tuneable parameter $\alpha$ allows the system to smoothly interpolate between homogeneous and heterogeneous activity distributions. 
 

Even though this linear approach successfully recovers heterogeneous activity distributions, capturing the diverse spectrum of social ties, from acquaintances to close friends, it suffers from a sociological limitation. In the preferential regime (e.g., for $\alpha \to -a_0$ in \cite{iniguez2023universal}), the connection probability of an alter grows indefinitely with its past activity, implying that an initial advantage creates an unbridgeable gap compared to other ties. This mechanism lacks a saturation effect: once a relationship is initiated, the ego focuses almost exclusively on that specific alter, continuously reinforcing the tie without limit. While this mimics an initial phase of intense curiosity or engagement, such as meeting an intriguing person at a social gathering, real-world human interactions exhibit different dynamics (Figure~\ref{fig:fig2}d). In actual social processes, after a transient period of high initial interest, curiosity naturally saturates, allowing the individual to explore new connections or eventually transition toward a more balanced, baseline interaction rate among established friends. The linear model fails to capture this saturation and exploration phase, treating social reinforcement as an unyielding, permanent bias.  

To address this limitation, we introduce a model that takes into account a saturation effect on the connection kernel via a parameter, that we call critical activity $a_c$. 
The new connection kernel reads
\begin{equation}\label{eq:pi_saturated}
\pi_{a_j} = \frac{q_{a_j}}{\sum_j^k{q_{a_j}}},
\end{equation}
where
\begin{equation}\label{eq:q}
q_{a_j} = 
\begin{cases} a_j +\alpha & \text{if } a \le a_c \\[2ex]
a_c + \alpha & \text{if } a > a_c.
\end{cases}
\end{equation}
This model relies on two key parameters: the random choice $\alpha$ of the original model, which retains the same regularizing meaning as described in \citet{iniguez2023universal}, and the critical activity $a_c$, which represents the saturation threshold of the connection kernel. To gain deeper insight into the model, it is informative to examine the asymptotic behaviour of the connection kernel $\pi_a$ under extreme regimes of these parameters:

\begin{itemize}
    \item $\alpha \to -a_0$, $a_c \to \infty$: In this regime, the critical activity is never reached, and the connection kernel simplifies to a form purely proportional to past interactions, i.e., $\pi_a \propto a$. This limit maximizes the rich-get-richer effect, introducing a reinforcement mechanism that promotes a highly heterogeneous activity distribution.
    
    \item $\alpha \to -a_0$, $a_c > 0$: Here, the dynamics initially exhibits preferential attachment, leading to heterogeneous tie strengths. However, as the cumulative activity $a$ grows, multiple alters progressively reach the critical activity $a_c$. Beyond this point, the kernel flattens out, and the system asymptotically converges to a uniform scenario where alters are selected uniformly at random.
    
    \item $\alpha \to \infty$, $a_c \to \infty$: In this limit, the $\alpha$ parameter dominates and the historical dependency vanishes. The connection kernel converges to $\pi_a \approx 1/k$, meaning that the ego selects any of the $k$ alters with equal probability, regardless of past interaction history. This scenario leads to a completely homogeneous activity distribution across the network.
    
    \item $\alpha \to \infty$, $a_c \to 0$: This case reduces to the previous uniform scenario. When $a_c \to 0$, all alters are saturated directly, whereas for a large $\alpha$ further suppresses any residual historical bias, driving the network immediately toward a random choice dynamics.
\end{itemize}

In the presence of a finite critical activity ($a_c > 0$), the total number of events $\tau$ can be expressed as
\begin{equation}\label{eq:tau_saturated}
\tau = \sum_{a_j \in \{a \le a_c\}} a_j(\tau) + a_c \ n_{a \ge a_c}.
\end{equation}
The first term represents the sum of the activities over all alters whose activity remains below the critical activity, while the second is given by the product of the bandwidth itself and the number of alters who have already reached it ($n_{a \ge a_c}$). It is worth noting that all nodes are expected to reach bandwidth at $\tau_c = k a_c$.

Assuming $\alpha \to -a_0$ with a finite $a_c$, and defining $t'$ and $t''$ as the times at which the first and the last alters reach the critical activity, three distinct dynamical regimes can be identified:
\begin{itemize}
    \item $\tau \ll t'$: \textbf{Preferential Attachment (PA).} No alters have reached the critical activity yet; during this phase, the preferential attachment mechanism is established.
    \item $\tau \approx t''$: \textbf{Stationarity.} Every alter has reached the critical activity. From this point onward, the system evolves by selecting alters uniformly at random.
    \item \textbf{Hybrid Regime.} At the intermediate stage between the two previous regimes, a hybrid phase emerges where some, but not all, alters have reached the critical activity. In this case, there is no further increase in the probability of interacting with alters in the bandwidth, prompting the ego to explore the remaining neighbours.
\end{itemize}

An assumption of the model is that the number of alters is fixed at $k$. Although this neglects the formation of new ties or the dissolution of existing ones, it assumes the structural network changes occur at a sufficiently slower rate compared to the high communication frequency. Assuming a stable $k$ is thus a mathematically reasonable approximation for the observed period.

\label{subsec:fitmodel}
\subsection{Fitting of $\alpha$ and $a_c$}
As a first step we fit the parameter $\alpha$ of the simpler model introduced in \cite{iniguez2023universal}, which did not account for an activity bandwidth to represent a finite social capacity.
In this model, the $\alpha$ parameter controls how much an ego favours their already-active alters versus picking alters at random.
The probability $\pi_a$ that an alter with activity $a$ interacts with the ego at event time $\tau + 1$ is
\begin{equation}\label{eq:pa_old_model}
    \pi_a = \dfrac{a+\alpha}{\tau + k \alpha}.
\end{equation}
Introducing the preferentiality parameter $\beta = t_r / \alpha_r$, with $t_r = t - a_0$, $t$ representing the mean alter activity, and $\alpha_r = \alpha + a_0$, and solving the model analytically through a master equation, the resulting activity distribution reads (see \cite{iniguez2023universal} for the full derivation):
\begin{equation}
	p_a = p_0 \, \frac{a_r^{-1}}{\mathrm{B}(a_r, \alpha_r)} \left( 1 + \frac{1}{\beta} \right)^{-a_r}.
	\label{eq:pa}
\end{equation}

On this basis, we derive maximum likelihood estimates of the model parameter $\alpha$ for empirical egos drawn from both the online and offline datasets.
For each network we sample $100$ egos (or all of the egos in case of networks of smaller size), characterised by their degree $k$, minimum and maximum alter activity $a_0$ and $a_m$, and total and mean alter activity $\tau = \sum_i a_i$ and $t = \tau/k$ (See SI Section~S3 of \cite{iniguez2023universal}).
Assuming that the $k$ alter activities $\{a_i\}$ are independent and identically distributed random variables following $p_a$, the likelihood $L_\alpha$ that the sample $\{a_i\}$ is generated by Eq.~\eqref{eq:pa} for a given $\alpha$ satisfies
\begin{equation}
	d_\alpha \ln L_\alpha = k\left[ F_\alpha - \ln(1+\beta) \right],
	\label{eq:loglik}
\end{equation}
where $F_\alpha = \frac{1}{k}\sum_i \left[ \psi(a_r + \alpha_r) - \psi(\alpha_r) \right]$ is an average over all observed relative activities $a_r = a_i - a_0$ of the digamma function $\psi(\alpha) = d_\alpha \Gamma(\alpha)/\Gamma(\alpha)$, i.e.\ the logarithmic derivative of the gamma function $\Gamma(\alpha)$.
The $\alpha$ that maximizes $L_\alpha$ is then given implicitly by
\begin{equation}
	\alpha_r = \frac{t_r}{e^{F_\alpha} - 1},
	\label{eq:alpha}
\end{equation}
or, equivalently, by $\beta = e^{F_\alpha} - 1$.

To assess how plausibly the empirical data is described by the model activity distribution of Eq.~\eqref{eq:pa}, we perform a goodness-of-fit test based on the standard Kolmogorov--Smirnov statistic
\begin{equation}
	D = \max_{a_0 \leq a \leq a_m} |\Delta P_a|,
	\label{eq:ks}
\end{equation}
that is, the largest magnitude of the difference $\Delta P_a(t) = P_{\mathrm{data}}[a' \leq a] - P_a(t)$ between the cumulative distribution of alter activity in the data, $P_{\mathrm{data}}[a' \leq a]$, and that of the fitted model, $P_a(t) = \sum_{a' = a_0}^{a} p_{a'}(t)$, across all activities $a \in [a_0, a_m]$.
Given the sample $\{a_i\}$, we compute the estimate $\alpha$ numerically from Eq.~\eqref{eq:alpha} and the statistic $D$ from Eq.~\eqref{eq:ks}, with the model activity distribution following Eq.~\eqref{eq:pa}.
We then generate $n_{\mathrm{sim}} = 2500$ simulated activity samples $\{a_i\}_{\mathrm{sim}}$ from the model; for each one we obtain an associated estimate $\alpha_{\mathrm{sim}}$ and the corresponding statistic $D_{\mathrm{sim}}$.
The fraction of simulated statistics $D_{\mathrm{sim}}$ exceeding the data statistic $D$ defines the $p$-value of the goodness-of-fit test.
If this $p$-value is large enough ($p > 0.1$, with $0.1$ an arbitrary significance threshold), we do not reject the hypothesis that the model emulates the empirical data, and we treat the ego network as having a measurable preferentiality parameter $\beta$.
We check the robustness of our results with three other measures from the Cramér-von Mises family of test statistics:  the Cram\'er-von Mises ($W^2$), the Watson ($U^2$), and the Anderson-Darling ($A^2$) statistic.

Our goodness-of-fit test (Table~\ref{tab:table_stats_100}) shows that, excluding the pathological Primary School (coloc) case ($1\%$), $31$--$89\%$ of ego networks are well described by the model, with the fit varying by interaction type: face-to-face datasets are captured best (up to $89\%$, average $69\%$), followed by colocation ($59\%$) and online networks ($52\%$).

\paragraph{Fitting the social bandwidth $a_c$.}
The bandwidth $a_c$ is fit to each ego's empirically measured connection kernel. The saturated model of Eq.~\eqref{eq:pi_saturated} predicts a per-alter connection probability that grows linearly with activity up to the critical activity and flattens beyond it,
\begin{equation}
    \pi_a^{\mathrm{theo}}
    = \frac{\min(a, a_c) + \alpha}{\kappa},
    \qquad
    \kappa = \alpha\, k + \sum_{j=1}^{k} \min\!\big(a_j, a_c\big),
    \label{eq:kernel_theo}
\end{equation}
where the normalisation $\kappa$ ensures that $\sum_j \pi_{a_j}^{\mathrm{theo}} = 1$, and the $a_j$ are the ego's final alter activities.
We fit $a_c$ for each ego separately, holding $\alpha$ fixed at the maximum-likelihood value obtained above. Preferentiality $\beta = t_r/\alpha_r$ is determined entirely by $\alpha$ and the ego's mean and minimum activity through Eq.~\eqref{eq:alpha}, whereas $a_c$ is estimated solely from the shape of the connection kernel.
The bandwidth is chosen to minimise the root sum of squared errors between the theoretical and empirical kernels over the observed activity range,
\begin{equation}
    a_c = \arg\min_{a_c \ge 0}
    \sqrt{\sum_{a=a_0}^{a_m}
    \big(\pi_a^{\mathrm{theo}} - \pi_a^{\mathrm{emp}}\big)^2}.
    \label{eq:ac_objective}
\end{equation}
Because $a_c$ enters Eq.~\eqref{eq:kernel_theo} as a threshold, the objective in Eq.~\eqref{eq:ac_objective} is piecewise constant in $a_c$, with flat plateaus on which gradient-based methods and single-start searches stall. We therefore minimise it with the derivative-free Nelder--Mead algorithm, restarted from every integer cutoff in the range $[0, a_m]$ together with the heuristic seed $a_c = a_m/2$, and retain the solution with the lowest error. Applying this procedure to every ego yields the per-ego pairs $(\beta, a_c)$ analysed in Fig~\ref{fig:fig4}, where we restrict attention to egos with strength $\tau \ge 30$, for which the empirical kernel is sufficiently sampled.

\section*{Acknowledgments}
\label{sec:ack}
This work is the output of the workshop Complexity72h by Complexity Next Gen, held at Northeastern University London, London, UK, 22-26 June 2026 (\url{www.complexitynextgen.org/complexity72h/}).
We are grateful for data provision to 
Petter Holme (Email Kiel \& Uni, Forum, Messages, Dating). We acknowledge the Department of Computer Science of Aalto University for access to processed versions of the non-public datasets used here (Email Kiel \& Uni, Forum, Messages, Dating).
G.d.M. acknowledges funding from the postdoctoral fellowship of the Magnus Ehrnroothin s\"{a}\"{a}ti\"{o}.

\paragraph{Code availability.}
Code is publicly available at \url{https://github.com/gdm2019/offline_egonets_C72h}.

\bibliographystyle{unsrtnat}
\bibliography{references}

\clearpage

\appendix
\setcounter{table}{0} \setcounter{figure}{0} \renewcommand{\thetable}{S\arabic{table}} \renewcommand{\thefigure}{S\arabic{figure}}

\section{Appendix}
\subsection{Dataset Details}
\label{sec:data_code}

\newcommand{\grayrowrule}{%
\arrayrulecolor{black!16}\specialrule{0.25pt}{1.0pt}{1.0pt}\arrayrulecolor{black}}

\begingroup
\footnotesize
\setlength{\tabcolsep}{3.5pt}
\renewcommand{\arraystretch}{1.06}
\setlength{\LTcapwidth}{0.98\textwidth}

\begin{longtable}{
@{}
>{\RaggedRight\arraybackslash}p{0.22\textwidth}
>{\RaggedRight\arraybackslash}p{0.75\textwidth}
@{}
}

\toprule
\textbf{Dataset}\\ \textit{Recorded interaction} &
\textbf{Observation window / data used} \\
\midrule
\endfirsthead

\toprule
\textbf{Dataset and recorded interaction} &
\textbf{Observation window / data used} \\
\midrule
\endhead

\midrule
\multicolumn{2}{r}{\textit{Continued on next page}} \\
\endfoot

\endlastfoot
\textbf{Mobile (Wu 1--3)}~\citep{wu2010evidence} \newline
\emph{Short messages}
&
Short messages from a mobile phone operator, comprising three billing accounts from three separate companies collected over one month. Each event records anonymized sender and recipient mobile numbers with a hashed timestamp accurate to one second. The data are publicly available in the Supplementary Information of the original study.
\\
\grayrowrule

\textbf{Email (Enron)}~\citep{klimt2004enron,kunegis2013konect} \newline
\emph{Emails}
&
Email communication from the Enron Corporation spanning 1999--2003. The corpus was made public following legal action by the Federal Energy Regulatory Commission in the US. We use the Koblenz network collection, containing emails among internal and external Enron addresses, and remove events with identical sender and recipient. Data are publicly available from KONECT.
\\
\grayrowrule

\textbf{Email (Kiel)}~\citep{ebel2002scale,saramaki2015exploring} \newline
\emph{Emails}
&
Log files from the email server at Kiel University, recording the source and destination of every email sent to or from student accounts over 112 days. See Section~\ref{sec:ack} for data acknowledgment.
\\
\grayrowrule

\textbf{Email (Uni)}~\citep{saramaki2015exploring,eckmann2004entropy} \newline
\emph{Emails}
&
Log files from one of the main mail servers at an unnamed university, covering 83 days. The data are restricted to internal mail within the institution. The number of recorded events differs slightly from the value reported in the original publication when calculated directly from the available data. See Section~\ref{sec:ack} for data acknowledgment.
\\
\grayrowrule

\textbf{Email (EU)}~\citep{leskovec2007graph,paranjape2017motifs} \newline
\emph{Emails}
&
Email communication within a large European research institution from October 2003 to May 2005. We focus only on institution members and on emails exchanged between them. Data are publicly available from SNAP.
\\
\grayrowrule

\textbf{Facebook}~\citep{saramaki2015exploring,viswanath2009evolution} \newline
\emph{Online messages}
&
Wall-post interactions from a large subset of the Facebook New Orleans social network. Facebook links were crawled in late 2008 and early 2009, and wall-post activity spans September 2006 to January 2009. Data are publicly available from the original Facebook New Orleans data release.
\\
\grayrowrule

\textbf{Messages}~\citep{saramaki2015exploring,said2010social,karimi2014structural} \newline
\emph{Online messages}
&
Private user-to-user messages from Filmtipset, a Swedish social movie recommendation community active since 2000. The available data cover time-stamped communication events over seven years. See Section~\ref{sec:ack} for data acknowledgment.
\\
\grayrowrule

\textbf{Forum}~\citep{saramaki2015exploring,said2010social,karimi2014structural} \newline
\emph{Online forum comments}
&
Forum interactions from Filmtipset, where users comment on posts written by other users. These data come from the same online community as the Messages dataset but represent open forum communication rather than private messages. See Section~\ref{sec:ack} for data acknowledgment.
\\
\grayrowrule

\textbf{Dating}~\citep{saramaki2015exploring,holme2004structure} \newline
\emph{Online community interactions}
&
Activity from pussokram.com, a Swedish online community primarily intended for romantic communication and targeted at adolescents and young adults. The data cover 512 days and include private messages, guestbook signing, friendship requests, and friendships. See Section~\ref{sec:ack} for data acknowledgment.
\\
\grayrowrule

\textbf{College}~\citep{opsahl2009clustering,panzarasa2009patterns} \newline
\emph{Online messages}
&
Private messages sent on a Facebook-like online social network for students at the University of California, Irvine, from April to October 2004. Users could search for others and initiate conversations based on profile information. The dataset is publicly available from Tore Opsahl's dataset collection and from SNAP.
\\

\bottomrule
\\
\caption{\textbf{Online communication datasets used in the analysis.}
For each online dataset, the table reports the recorded interaction type and the observation window or subset used in the analysis. Source references are shown next to the dataset names.}
\label{tab:online_data_sources}

\end{longtable}
\endgroup

\begingroup
\footnotesize
\setlength{\tabcolsep}{3.5pt}
\renewcommand{\arraystretch}{1.07}
\setlength{\LTcapwidth}{0.98\textwidth}

\begin{longtable}{
@{}
>{\RaggedRight\arraybackslash}p{0.195\textwidth}
>{\RaggedRight\arraybackslash}p{0.195\textwidth}
>{\RaggedRight\arraybackslash}p{0.56\textwidth}
@{}
}

\toprule
\textbf{Dataset family and source} &
\textbf{Dataset representation used} &
\textbf{Observation window / data used} \\
\midrule
\endfirsthead

\toprule
\textbf{Dataset family and source} &
\textbf{Dataset representation used} &
\textbf{Observation window / data used} \\
\midrule
\endhead

\midrule
\multicolumn{3}{r}{\textit{Continued on next page}} \\
\endfoot

\endlastfoot

\textbf{Hospital}~\citep{genois2018can,vanhems2013estimating}
&
\emph{Colocation (RFID)}\newline
\emph{Face-to-face (RFID)}
&
Physical proximity in the geriatric unit of a university hospital in Lyon, France, collected over 4 days in 2010 using the SocioPatterns protocol. Contacts were recorded among healthcare workers, among patients, and between patients and healthcare workers. The analysis uses the corresponding colocation and face-to-face representations.
\\
\grayrowrule

\textbf{Office}~\citep{genois2018can}
&
\emph{Colocation (RFID), 2013}\newline
\emph{Colocation (RFID), 2015}\newline
\emph{Face-to-face (RFID), 2013}
&
Physical proximity within an office building in France. The 2013 and 2015 colocation datasets were measured over two weeks and split into working-day subsets; we use the first working day for 2013 and the seventh working day for 2015. The 2013 face-to-face dataset was split into two weekly subsets, and we use the first week.
\\
\grayrowrule

\textbf{Primary School (Lyon)}~\citep{stehle2011high}
&
\emph{Colocation (RFID)}\newline
\emph{Face-to-face (RFID)}
&
Physical proximity between students and teachers at a primary school in Lyon, France, collected over two consecutive days in 2009 using wearable RFID sensors. We split the dataset into daily subsets. We use the first day for the colocation representation and the second day for the Face-to-Face one.
\\
\grayrowrule

\textbf{High School}~\citep{mastrandrea2015contact,genois2018can}
&
\emph{Face-to-face (RFID), 2011}\newline
\emph{Face-to-face (RFID), 2012}\newline
\emph{Face-to-face (RFID), 2013}\newline
\emph{Colocation (RFID), 2013}
&
Physical proximity within high schools in Marseilles, France. The face-to-face datasets include interactions among students recorded in 2011, 2012, and 2013 using the SocioPatterns protocol. We use the first day for 2011, the first week for 2012, and the second day for 2013. The broad-proximity/colocation 2013 dataset was split into five school days, and we use the third day.
\\
\grayrowrule

\textbf{University (MIT)}~\citep{eagle2006reality}
&
\emph{Bluetooth proximity}
&
Physical proximity between students at the Massachusetts Institute of Technology, measured by mobile-phone Bluetooth handshakes from September 2004 to May 2005. Contacts are recorded with a temporal resolution of 5 minutes and an approximate spatial range of 7--10 meters.
\\
\grayrowrule

\textbf{University (UC San Diego)}~\citep{mcnett2005access,chaintreau2007impact}
&
\emph{Wi-Fi proximity}
&
Physical proximity among freshman students at U.C. San Diego in 2002, inferred from wireless network availability traces collected from HP Jornada devices used as personal digital assistants. We use the first 3 days of data.
\\
\grayrowrule

\textbf{Conference (SFHH)}~\citep{genois2018can}
&
\emph{Colocation (RFID)}
&
Proximity/colocation contacts collected during a scientific conference in Nice, France, over 2 days in 2009. We split the dataset into daily subsets and use the first day.
\\
\grayrowrule

\textbf{Conference (hypertext)}~\citep{isella2011s}
&
\emph{Face-to-face (RFID)}
&
Face-to-face proximity among attendees of the ACM Hypertext 2009 conference, recorded over 2.5 days using wearable radio badges. We split the dataset into three daily subsets and use the second day.
\\
\grayrowrule

\textbf{Exhibition}~\citep{isella2011s}
&
\emph{Face-to-face (RFID)}
&
Face-to-face proximity collected during the \emph{INFECTIOUS: STAY AWAY} art-science exhibition at the Science Gallery in Dublin, Ireland, in 2009, measured using the SocioPatterns protocol.
\\

\bottomrule
\\
\caption{\textbf{Offline colocation and face-to-face datasets used in the analysis.}
For each offline dataset family, the table reports the interaction representation used in the analysis and the observation window or selected subset. RFID datasets follow the SocioPatterns protocol unless otherwise stated; this protocol records close-range proximity with a temporal resolution of 20 seconds and an approximate spatial range of 1.5 meters. All offline datasets are available through the Netzschleuder open repository~\citep{peixoto2020netzschleuder}.}
\label{tab:offline_data_sources}

\end{longtable}
\endgroup

\subsection{Summary statistics and distributional checks}
\label{app:summary_statistics}

Table~\ref{tab:basic_properties_iet} summarizes the online communication, offline face-to-face, and offline colocation/proximity datasets used in the analysis. For each dataset, we report the unfiltered number of individuals $N_u$, the total number of recorded events $V$, and the number of retained egos $N$ after applying the filter $t>a_0$. For retained egos, we report the average degree $\langle k\rangle$, average strength $\langle \tau\rangle$, average mean alter activity $\langle t\rangle$, average minimum alter activity $\langle a_0\rangle$, and average maximum alter activity $\langle a_m\rangle$. These quantities summarize the size of ego networks, the volume of repeated interaction, and the contrast between weak and strong ego--alter ties. We also report the raw mean edge-level inter-event time $\langle \mathrm{IET}\rangle$ and the burstiness index $B$.

\begin{table*}[!htbp]
\centering
\small
\setlength{\tabcolsep}{4.2pt}
\renewcommand{\arraystretch}{1.12}
\resizebox{\textwidth}{!}{%
\begin{tabular}{llrrrrrrrrrr}
\toprule
\textbf{Dataset} & \textbf{Event} & $N_u$ & $V$ & $N$ & $\langle k \rangle$ & $\langle \tau \rangle$ & $\langle t \rangle$ & $\langle a_0 \rangle$ & $\langle a_m \rangle$ & $\langle \mathrm{IET} \rangle$ & $B$ \\
\midrule
\multicolumn{12}{l}{\textbf{Online datasets}} \\
\hline
Mobile (Wu 1) & Short messages & 44,090 & 544,817 & 16,050 & 4.55 & 52.93 & 12.74 & 1.84 & 38.10 & 428915.09 & 0.58 \\
Mobile (Wu 2) & Short messages & 71,042 & 636,629 & 20,534 & 4.71 & 43.86 & 10.66 & 1.91 & 29.86 & 498571.69 & 0.55 \\
Mobile (Wu 3) & Short messages & 14,273 & 140,611 & 4,215 & 6.27 & 52.72 & 10.66 & 1.79 & 33.29 & 527300.64 & 0.53 \\
Email (Enron) & Emails & 86,978 & 1,134,990 & 21,984 & 22.52 & 96.43 & 3.26 & 1.15 & 16.27 & 800722.82 & 0.48 \\
Email (Kiel) & Emails & 57,189 & 431,864 & 9,842 & 13.05 & 65.68 & 5.99 & 1.79 & 25.86 & 221485.32 & 0.43 \\
Email (Uni) & Emails & 3,188 & 308,730 & 2,456 & 25.49 & 250.10 & 9.14 & 1.12 & 61.18 & 182188.28 & 0.38 \\
Email (EU) & Emails & 986 & 332,334 & 866 & 36.92 & 766.99 & 18.22 & 1.08 & 194.64 & 615529.45 & 0.56 \\
Facebook & Online messages & 45,813 & 854,612 & 31,429 & 11.04 & 53.22 & 4.08 & 1.17 & 17.07 & 1728757.79 & 0.47 \\
Messages & Online messages & 35,623 & 478,015 & 20,252 & 8.37 & 45.14 & 3.84 & 1.23 & 17.85 & 1107801.44 & 0.67 \\
Dating & Online messages & 28,972 & 430,826 & 16,239 & 13.05 & 51.44 & 3.44 & 1.11 & 13.29 & 289190.85 & 0.61 \\
Forum & Online messages & 7,084 & 1,428,493 & 4,122 & 65.22 & 691.15 & 2.83 & 1.01 & 57.41 & 966614.23 & 0.66 \\
College & Online messages & 1,899 & 59,835 & 1,303 & 20.48 & 90.90 & 3.62 & 1.07 & 17.22 & 104142.74 & 0.67 \\
\midrule
\multicolumn{12}{l}{\textbf{Offline colocation \& face-to-face datasets}} \\
\hline
Hospital & Colocation (RFID) & 66 & 51,012 & 60 & 16.80 & 1669.98 & 124.58 & 13.90 & 460.87 & 205.78 & 0.65 \\
Office (2013) & Colocation (RFID) & 94 & 93,507 & 91 & 23.70 & 2054.21 & 85.82 & 12.98 & 506.01 & 337.48 & 0.65 \\
Office (2015) & Colocation (RFID) & 218 & 352,254 & 216 & 53.36 & 3260.98 & 59.33 & 10.72 & 461.00 & 257.83 & 0.60 \\
Primary School (Lyon) & Colocation (RFID) & 237 & 3,503,893 & 237 & 182.48 & 29568.72 & 160.98 & 12.73 & 721.82 & 111.69 & 0.76 \\
High School (2013) & Colocation (RFID) & 327 & 3,688,551 & 325 & 70.17 & 22698.67 & 321.07 & 12.80 & 1265.21 & 75.34 & 0.65 \\
University (MIT) & Colocation(Bluetooth) & 94 & 1,083,906 & 94 & 35.40 & 23061.83 & 609.52 & 16.22 & 5391.28 & 50319.85 & 0.74 \\
Conference (SFHH) & Colocation (RFID) & 396 & 794,208 & 396 & 146.97 & 4011.15 & 24.27 & 10.01 & 141.21 & 596.76 & 0.56 \\
University (UC San Diego) & Colocation(Wi-Fi) & 67 & 24,676 & 53 & 31.17 & 923.47 & 27.43 & 11.26 & 79.79 & 6305.53 & 0.51 \\
Hospital & Face-to-face (RFID) & 72 & 30,065 & 70 & 14.40 & 858.16 & 47.84 & 11.49 & 210.49 & 1841.80 & 0.76 \\
Office (2013) & Face-to-face (RFID) & 70 & 3,599 & 58 & 3.72 & 119.07 & 31.31 & 14.40 & 58.86 & 4196.97 & 0.69 \\
Primary School (Lyon) & Face-to-face (RFID) & 237 & 53,225 & 233 & 13.21 & 456.35 & 33.66 & 11.06 & 111.15 & 503.33 & 0.56 \\
High School (2011) & Face-to-face (RFID) & 103 & 8,556 & 75 & 4.67 & 193.96 & 51.24 & 16.65 & 107.99 & 265.56 & 0.69 \\
High School (2012) & Face-to-face (RFID) & 169 & 28,852 & 155 & 6.14 & 368.72 & 59.80 & 17.45 & 176.03 & 3024.28 & 0.73 \\
High School (2013) & Face-to-face (RFID) & 300 & 42,215 & 250 & 4.74 & 329.95 & 74.47 & 19.66 & 188.78 & 239.09 & 0.67 \\
Conference (hypertext) & Face-to-face (RFID) & 91 & 5,083 & 68 & 4.69 & 138.75 & 31.43 & 17.04 & 53.93 & 417.92 & 0.67 \\
Exhibition & Face-to-face (RFID) & 9,074 & 338,995 & 4,755 & 3.17 & 102.78 & 34.45 & 21.90 & 49.50 & 72.07 & 0.38 \\
\bottomrule
\end{tabular}%
}
\caption{\small \textbf{Summary statistics of online communication, offline face-to-face, and offline colocation datasets.} For each dataset, $N_u$ denotes the unfiltered number of individuals, $V$ the number of recorded events, and $N$ the number of retained egos after applying the filter $t>a_0$, where $t=\tau/k$ is the mean alter activity and $a_0$ is the minimum alter activity of the ego. For retained egos, we report the average degree $\langle k\rangle$, average strength $\langle \tau\rangle$, average mean alter activity $\langle t\rangle$, average minimum alter activity $\langle a_0\rangle$, and average maximum alter activity $\langle a_m\rangle$. We also report the raw mean edge-level inter-event time $\langle \mathrm{IET}\rangle$, computed between repeated events of the same ego--alter pair in the original timestamp units of each dataset, and the unitless burstiness index $B$, where larger positive values indicate more temporally clustered repeated interactions.}
\label{tab:basic_properties_iet}
\end{table*}

\begin{figure*}
    \centering
    \includegraphics[width=0.99\textwidth]{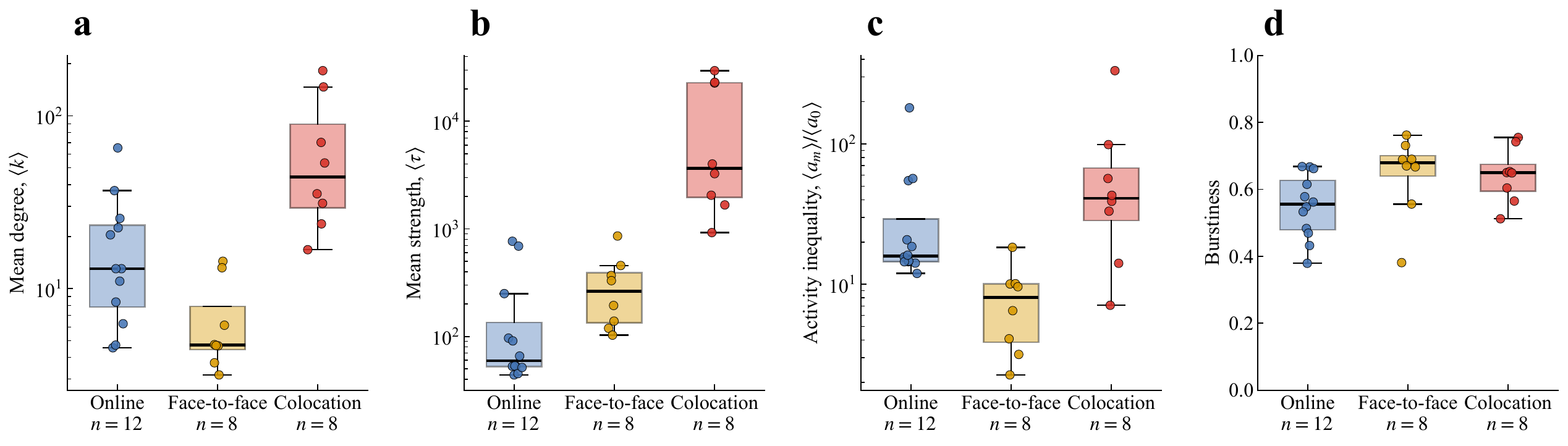}
    \caption{\small \textbf{Dataset-level summary of ego network statistics by measurement group.}
Each point represents one dataset, and box plots summarize variation within each measurement group: online communication, offline face-to-face contact, and offline colocation/proximity. Panels a--d show mean degree $\langle k\rangle$, mean strength $\langle \tau\rangle$, activity inequality $\langle a_m\rangle/\langle a_0\rangle$, and edge-level inter-event-time burstiness $B$, respectively. Panels a--c use logarithmic scales.}
    \label{fig:appendix_summary_groups}
\end{figure*}

Figure~\ref{fig:appendix_summary_groups} provides a compact group-level comparison of selected statistics from Table~\ref{tab:basic_properties_iet}. Each point represents one dataset, and box plots summarize variation across online communication, offline face-to-face contact, and offline colocation/proximity datasets. Colocation/proximity datasets generally show higher mean degree $\langle k\rangle$ and mean strength $\langle \tau\rangle$ than face-to-face datasets, indicating that broader proximity measurements reconstruct larger and denser ego networks. This difference reflects measurement modality: face-to-face sensors capture close-range interactions, whereas colocation, Bluetooth, and Wi-Fi measurements capture broader shared presence. The activity-inequality ratio $\langle a_m\rangle/\langle a_0\rangle$ is above one in all groups, showing that ego activity is usually concentrated unevenly across alters.

\begin{figure*}[!htbp]
    \centering
    \includegraphics[
        width=0.99\textwidth,
        keepaspectratio
    ]{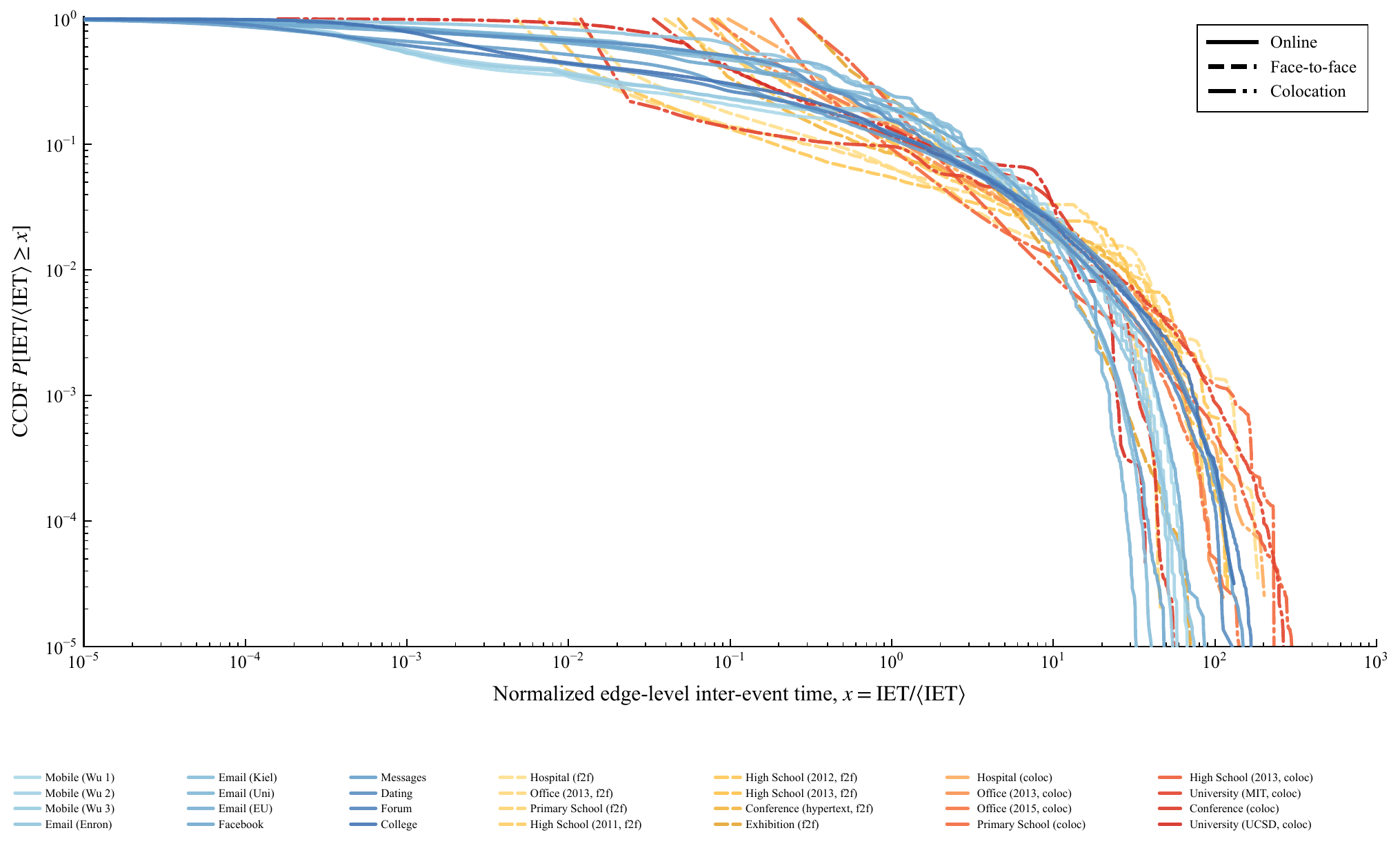}
    \caption{\textbf{Normalized edge-level inter-event-time distributions across online and offline datasets.}
    Complementary cumulative distribution functions (CCDFs) are shown for normalized edge-level inter-event times, $X=\mathrm{IET}/\langle \mathrm{IET}\rangle$, where IET is the time gap between consecutive repeated events of the same ego--alter pair. Line colour identifies individual datasets, while line style identifies measurement group: solid lines indicate online communication datasets, dashed lines indicate offline face-to-face contact datasets, and dash-dotted lines indicate offline colocation/proximity datasets. The broad right-tailed distributions indicate bursty repeated interactions across all measurement groups, consistent with the positive burstiness values reported in Table~\ref{tab:basic_properties_iet} and Figure~\ref{fig:appendix_summary_groups}d.}
    \label{fig:iet_combined}
\end{figure*}

Finally, Figure~\ref{fig:iet_combined} shows the normalized edge-level inter-event-time distributions, where each inter-event time is divided by the dataset-specific mean, $X=\mathrm{IET}/\langle \mathrm{IET}\rangle$. The broad right-tailed CCDFs indicate that repeated ego--alter interactions are not regularly spaced in time. Instead, many repeated events occur after short waiting times, while a smaller number are separated by much longer gaps. This is consistent with the positive burstiness values reported in Table~\ref{tab:basic_properties_iet} and summarized in Figure~\ref{fig:appendix_summary_groups}d. Together, these distributional checks support the main descriptive findings: ego network activity is heterogeneous, repeated interactions are temporally bursty, and the scale of reconstructed ego networks depends on the measurement modality.

\begin{figure*}[p]
    \centering
    \includegraphics[
        height=0.99\textheight,
        width=\textwidth,
        keepaspectratio
    ]{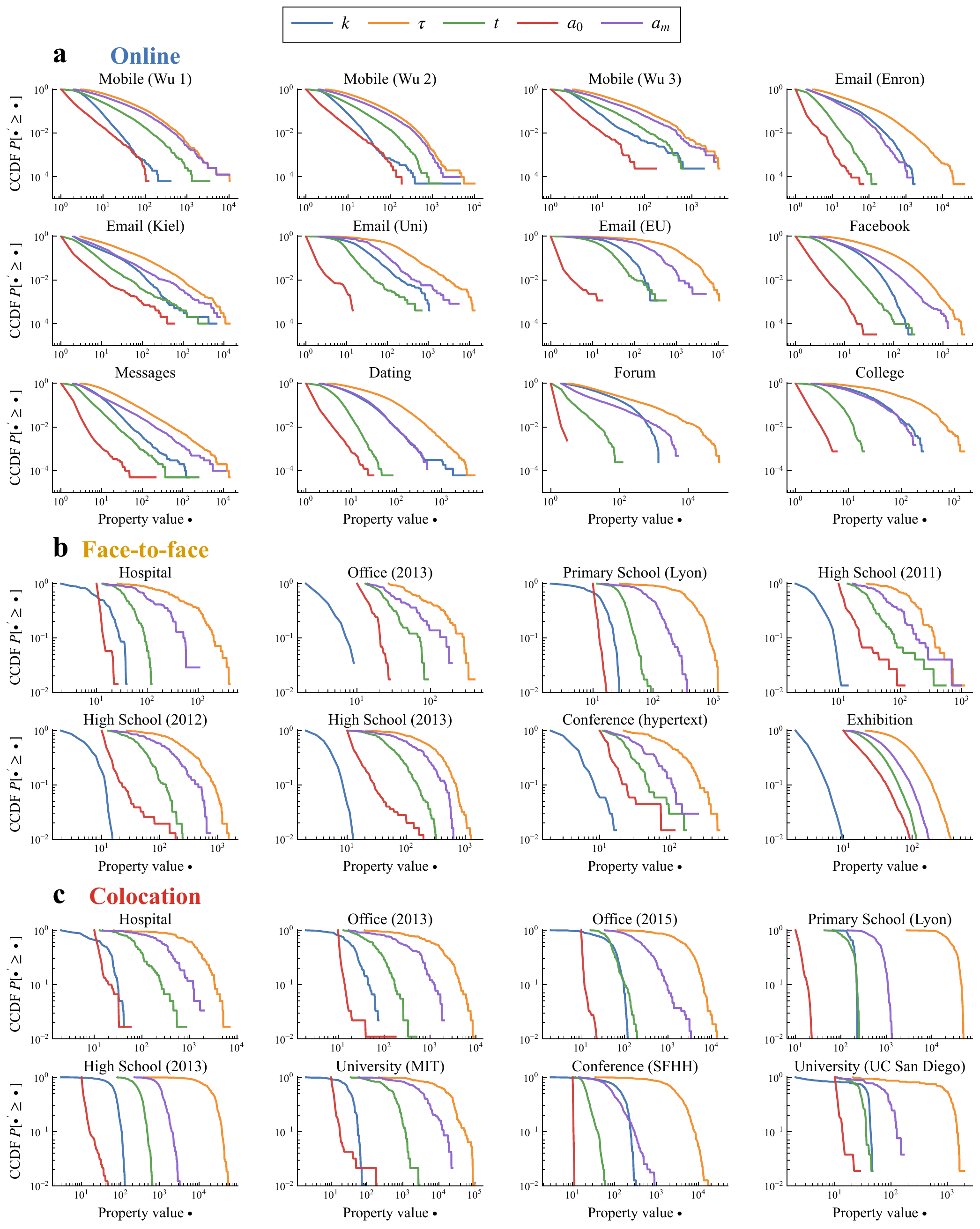}
    \caption{\small \textbf{Basic ego network property distributions across online and offline datasets.}
    Complementary cumulative distribution functions (CCDFs), $P[\bullet' \geq \bullet]$, are shown for degree $k$, strength $\tau$, mean alter activity $t=\tau/k$, minimum alter activity $a_0$, and maximum alter activity $a_m$. Datasets are grouped by measurement class: \textbf{a}, online communication datasets; \textbf{b}, offline face-to-face contact datasets; and \textbf{c}, offline colocation datasets. Each subplot corresponds to one dataset. The distributions show that ego network size and repeated activity are highly heterogeneous within datasets, and that colocation measurements generally reconstruct larger and denser ego networks than face-to-face measurements.}
    \label{fig:basic_properties_three_blocks}
\end{figure*}

Figure~\ref{fig:basic_properties_three_blocks} complements these averages by showing the full distributions of ego network properties for every dataset. The complementary cumulative distribution functions (CCDFs) of degree $k$, strength $\tau$, mean alter activity $t=\tau/k$, minimum alter activity $a_0$, and maximum alter activity $a_m$ span several orders of magnitude. This confirms that average values alone mask substantial within-dataset variation: most egos have relatively low activity, whereas a smaller number of egos have many alters or repeated interactions. The three measurement groups also show distinct patterns. Online datasets tend to have broad-tailed distributions, face-to-face datasets are generally more compact, and colocation/proximity datasets are often shifted toward larger values, especially for $\tau$, $t$, and $a_m$.

\begin{figure*}[!htbp]
    \centering
    \includegraphics[width=\linewidth]{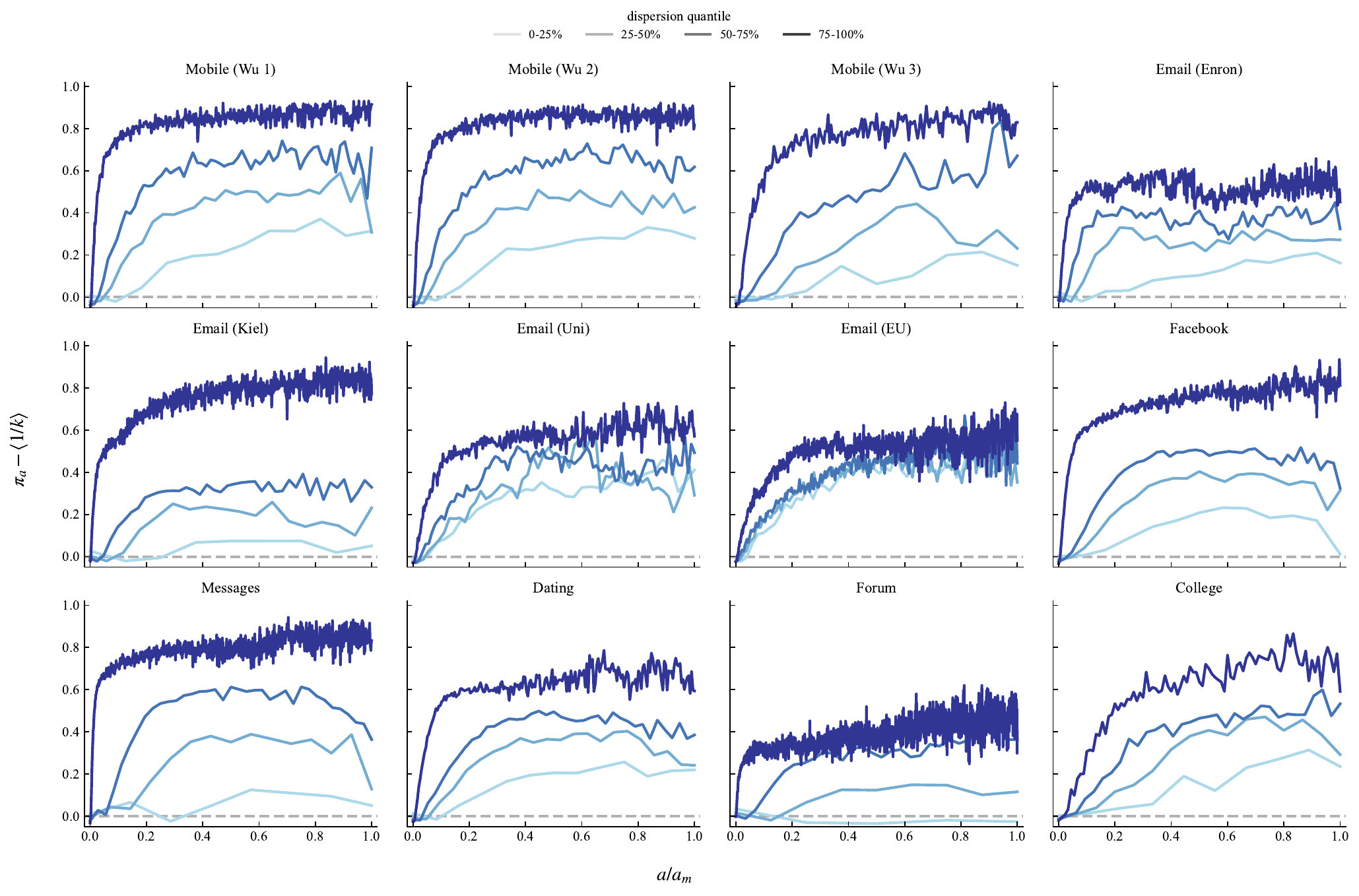}
    \caption{\textbf{Connection kernel by dispersion for each online dataset.} Relative connection kernel $\pi_a - \langle 1/k \rangle$ as a function of normalised alter activity $a/a_m$ aggregated over egos of online networks in the same dispersion quartile.      
    On the x-axis, activity levels are normalised to $[0,1]$ by the maximum activity level $a_m$. The baseline at $0$ denotes the case where communication events are distributed randomly.}
    \label{fig:each_kernel_online}
\end{figure*}

\begin{figure*}[!htbp]
    \centering
    \includegraphics[width=\linewidth]{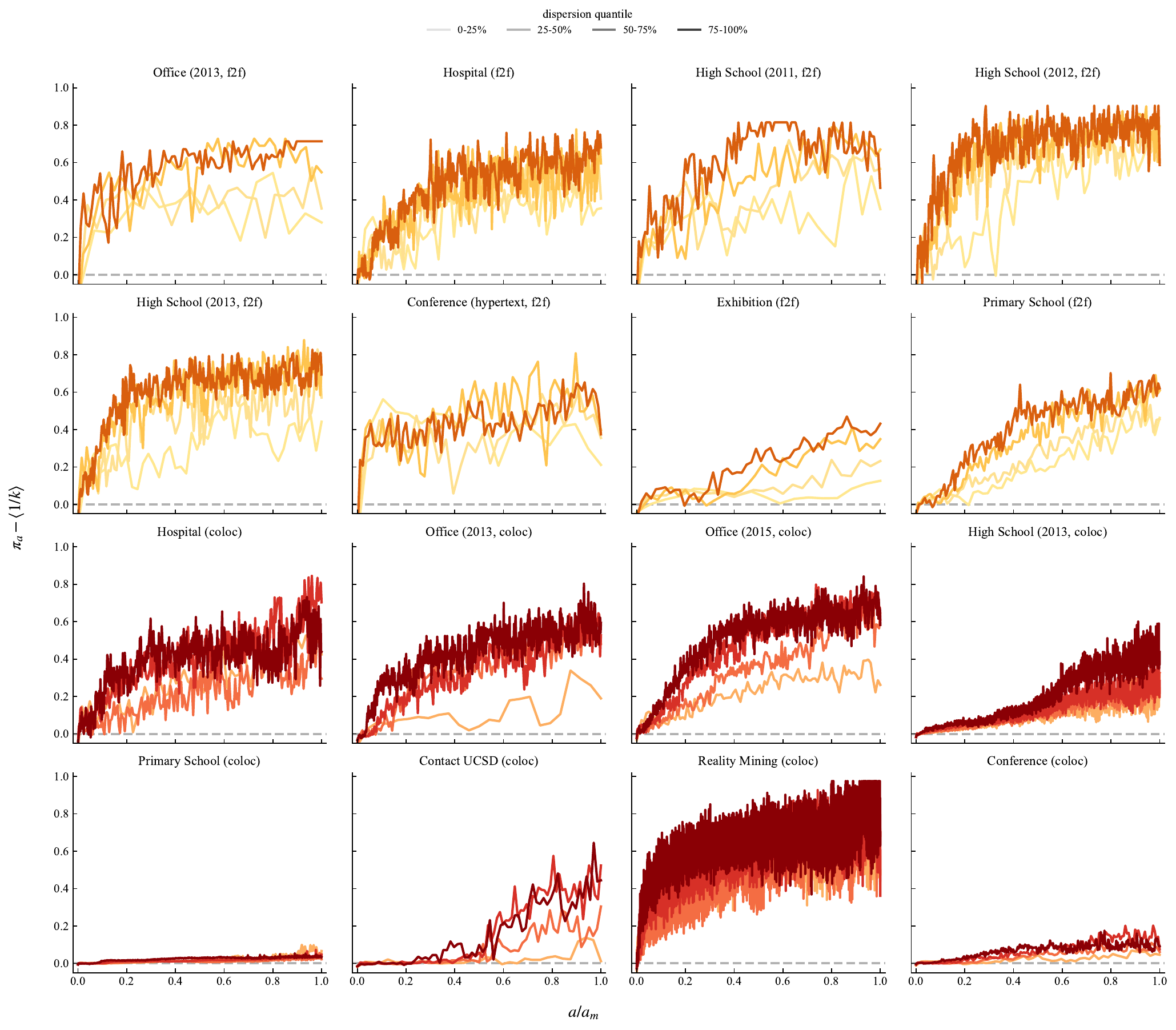}
    \caption{\textbf{Connection kernel by dispersion for each offline dataset.} Relative connection kernel $\pi_a - \langle 1/k \rangle$ as a function of normalised alter activity $a/a_m$ aggregated over egos of offline (face-to-face and colocation) networks in the same dispersion quartile.      
    On the x-axis, activity levels are normalised to $[0,1]$ by the maximum activity level $a_m$. The baseline at $0$ denotes the case where communication events are distributed randomly.}
    \label{fig:each_kernel_offline}
\end{figure*}

\begin{figure*}[!htbp]
    \centering
    \includegraphics[width=\linewidth]{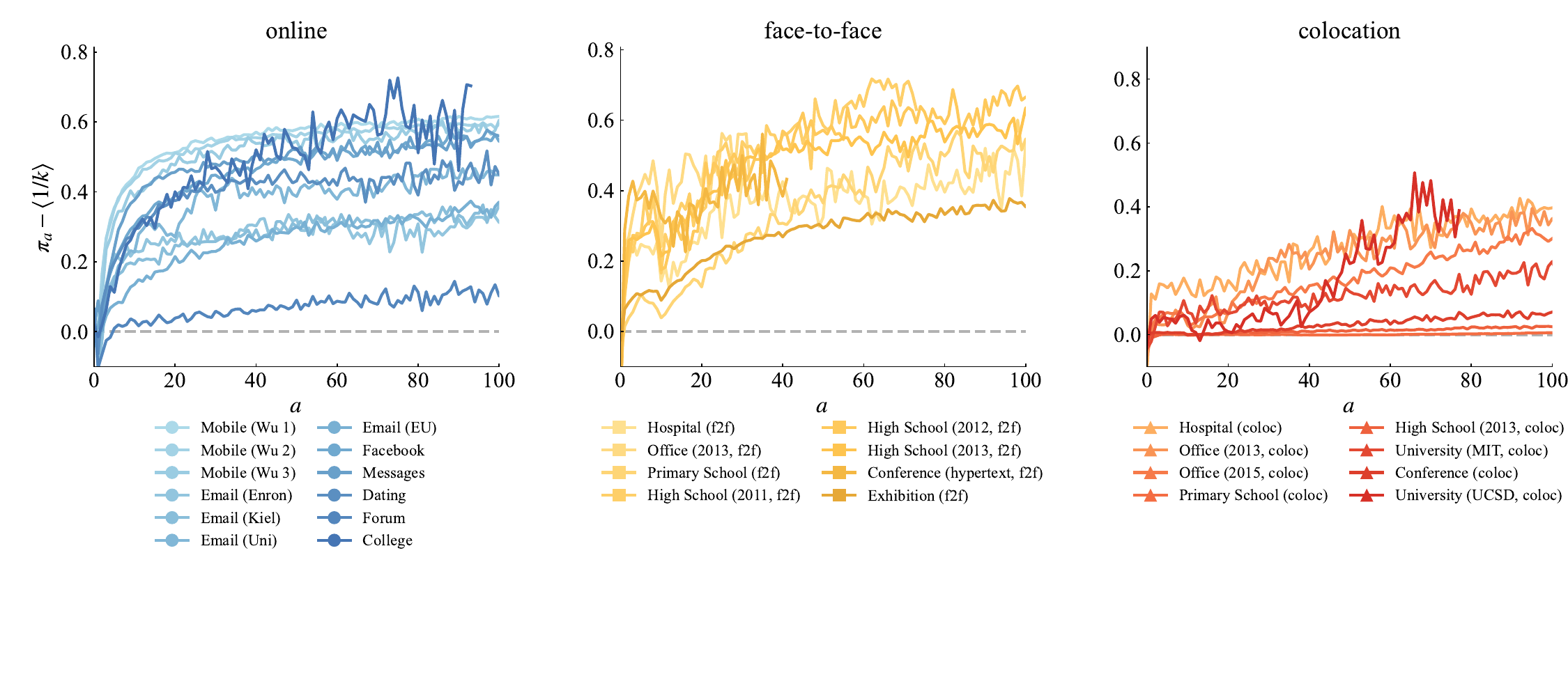}
    \caption{\textbf{Relative connection kernel for all datasets.} Relative connection kernel $\pi_a - \langle 1/k \rangle$ as a function of normalised alter activity $a/a_m$ aggregated over all egos of online (left), face-toface (middle) and colocation (right) datasets.      
    On the x-axis, activity levels are normalised to $[0,1]$ by the maximum activity level $a_m$. The baseline at $0$ denotes the case where communication events are distributed randomly.}
    \label{fig:kernel_all_egos}
\end{figure*}

\end{document}